\documentclass[aps,prd, floatfix, twocolumn,superscriptaddress, nofootinbib, preprintnumbers]{revtex4-2}
\pdfoutput=1
\usepackage{aas_macros,natbib}
\usepackage{graphicx}
\usepackage{color}
\usepackage[hidelinks]{hyperref}
\hypersetup{colorlinks=true,linkcolor=blue,citecolor=blue,urlcolor=blue}
\usepackage{float}
\usepackage{graphicx}
\usepackage{makecell}
\usepackage{mathtools, bm}
\usepackage{orcidlink}
\usepackage[usestackEOL]{stackengine}
\usepackage[format=plain,labelfont=bf,up,]{caption}
\usepackage{subcaption}

\begin{document}
\preprint{DES-2022-0684}
\preprint{FERMILAB-PUB-23-239-T}

\title{The Kinematic Sunyaev-Zel'dovich Effect with ACT, DES, and BOSS: a Novel Hybrid Estimator}

\author{M.~Mallaby-Kay \orcidlink{0000-0002-2018-3807}}
\affiliation{Department of Astronomy and Astrophysics, University of Chicago, Chicago, IL 60637, USA}

\author{S.~Amodeo \orcidlink{0000-0002-4200-9965}}
\affiliation{Universit\'e de Strasbourg, CNRS, Observatoire astronomique de Strasbourg, UMR 7550, F-67000 Strasbourg, France}

\author{J.~C.~Hill \orcidlink{0000-0002-9539-0835}}
\affiliation{Department of Physics, Columbia University, New York, NY 10027, USA}

\author{M.~Aguena}
\affiliation{Laborat\'orio Interinstitucional de e-Astronomia - LIneA, Rua Gal. Jos\'e Cristino 77, Rio de Janeiro, RJ - 20921-400, Brazil}

\author{S.~Allam}
\affiliation{Fermi National Accelerator Laboratory, P. O. Box 500, Batavia, IL 60510, USA}

\author{O.~Alves}
\affiliation{Department of Physics, University of Michigan, Ann Arbor, MI 48109, USA}

\author{J.~Annis}
\affiliation{Fermi National Accelerator Laboratory, P. O. Box 500, Batavia, IL 60510, USA}

\author{N.~Battaglia \orcidlink{0000-0001-5846-0411}}
\affiliation{Department of Astronomy, Cornell University, Ithaca, NY 14853, USA}

\author{E.~S.~Battistelli \orcidlink{0000-0001-5210-7625}}
\affiliation{Physics Department, Sapienza University of Rome,
Piazzale Aldo Moro 5, I-00185 Rome, Italy}

\author{E.~J.~Baxter}
\affiliation{Institute for Astronomy, University of Hawai'i, 2680 Woodlawn Drive, Honolulu, HI 96822, USA}

\author{K.~Bechtol}
\affiliation{Physics Department, 2320 Chamberlin Hall, University of Wisconsin-Madison, 1150 University Avenue Madison, WI  53706-1390}

\author{M.~R.~Becker}
\affiliation{Argonne National Laboratory, 9700 South Cass Avenue, Lemont, IL 60439, USA}

\author{E.~Bertin}
\affiliation{CNRS, UMR 7095, Institut d'Astrophysique de Paris, F-75014, Paris, France}
\affiliation{Sorbonne Universit\'es, UPMC Univ Paris 06, UMR 7095, Institut d'Astrophysique de Paris, F-75014, Paris, France}

\author{J.~R.~Bond}
\affiliation{Canadian Institute for Theoretical Astrophysics, University of Toronto, 60 St. George St., Toronto, ON M5S 3H4, Canada}

\author{D.~Brooks}
\affiliation{Department of Physics \& Astronomy, University College London, Gower Street, London, WC1E 6BT, UK}

\author{E.~Calabrese \orcidlink{0000-0003-0837-0068}}
\affiliation{School of Physics and Astronomy, Cardiff University, The Parade, Cardiff, Wales CF24 3AA, UK}

\author{A.~Carnero~Rosell}
\affiliation{Instituto de Astrofisica de Canarias, E-38205 La Laguna, Tenerife, Spain}
\affiliation{Laborat\'orio Interinstitucional de e-Astronomia - LIneA, Rua Gal. Jos\'e Cristino 77, Rio de Janeiro, RJ - 20921-400, Brazil}
\affiliation{Universidad de La Laguna, Dpto. Astrofisica, E-38206 La Laguna, Tenerife, Spain}

\author{M.~Carrasco~Kind}
\affiliation{Center for Astrophysical Surveys, National Center for Supercomputing Applications, 1205 West Clark St., Urbana, IL 61801, USA}
\affiliation{Department of Astronomy, University of Illinois at Urbana-Champaign, 1002 W. Green Street, Urbana, IL 61801, USA}

\author{J.~Carretero}
\affiliation{Institut de F\'{\i}sica d'Altes Energies (IFAE), The Barcelona Institute of Science and Technology, Campus UAB, 08193 Bellaterra (Barcelona) Spain}

\author{A.~Choi}
\affiliation{NASA Goddard Space Flight Center, 8800 Greenbelt Rd, Greenbelt, MD 20771, USA}

\author{M.~Crocce}
\affiliation{Institut d'Estudis Espacials de Catalunya (IEEC), 08034 Barcelona, Spain}
\affiliation{Institute of Space Sciences (ICE, CSIC),  Campus UAB, Carrer de Can Magrans, s/n,  08193 Barcelona, Spain}

\author{L.~N.~da Costa}
\affiliation{Laborat\'orio Interinstitucional de e-Astronomia - LIneA, Rua Gal. Jos\'e Cristino 77, Rio de Janeiro, RJ - 20921-400, Brazil}

\author{M.~E.~S.~Pereira}
\affiliation{Hamburger Sternwarte, Universit\"{a}t Hamburg, Gojenbergsweg 112, 21029 Hamburg, Germany}

\author{J.~De~Vicente}
\affiliation{Centro de Investigaciones Energ\'eticas, Medioambientales y Tecnol\'ogicas (CIEMAT), Madrid, Spain}

\author{S.~Desai}
\affiliation{Department of Physics, IIT Hyderabad, Kandi, Telangana 502285, India}

\author{J.~P.~Dietrich}
\affiliation{University Observatory, Faculty of Physics, Ludwig-Maximilians-Universit\"at, Scheinerstr. 1, 81679 Munich, Germany}

\author{P.~Doel}
\affiliation{Department of Physics \& Astronomy, University College London, Gower Street, London, WC1E 6BT, UK}

\author{C.~Doux}
\affiliation{Department of Physics and Astronomy, University of Pennsylvania, Philadelphia, PA 19104, USA}
\affiliation{Universit\'e Grenoble Alpes, CNRS, LPSC-IN2P3, 38000 Grenoble, France}

\author{A.~Drlica-Wagner}
\affiliation{Department of Astronomy and Astrophysics, University of Chicago, Chicago, IL 60637, USA}
\affiliation{Fermi National Accelerator Laboratory, P. O. Box 500, Batavia, IL 60510, USA}
\affiliation{Kavli Institute for Cosmological Physics, University of Chicago, Chicago, IL 60637, USA}

\author{J.~Dunkley \orcidlink{0000-0002-7450-2586}}
\affiliation{Joseph Henry Laboratories of Physics, Jadwin Hall, Princeton University, Princeton, NJ, USA 08544}
\affiliation{Department of Astrophysical Sciences, Peyton Hall, Princeton University, Princeton, NJ USA 08544}

\author{J.~Elvin-Poole}
\affiliation{Department of Physics and Astronomy, University of Waterloo, 200 University Ave W, Waterloo, ON N2L 3G1, Canada}

\author{S.~Everett}
\affiliation{Jet Propulsion Laboratory, California Institute of Technology, 4800 Oak Grove Dr., Pasadena, CA 91109, USA}

\author{S.~Ferraro \orcidlink{0000-0003-4992-7854}}
\affiliation{Lawrence Berkeley National Laboratory, One Cyclotron Road, Berkeley, CA 94720, USA}

\author{I.~Ferrero}
\affiliation{Institute of Theoretical Astrophysics, University of Oslo. P.O. Box 1029 Blindern, NO-0315 Oslo, Norway}

\author{J.~Frieman}
\affiliation{Fermi National Accelerator Laboratory, P. O. Box 500, Batavia, IL 60510, USA}
\affiliation{Kavli Institute for Cosmological Physics, University of Chicago, Chicago, IL 60637, USA}

\author{P.~A.~Gallardo \orcidlink{0000-0001-9731-3617}}
\affiliation{Kavli Institute for Cosmological Physics, University of Chicago, Chicago, IL 60637, USA}

\author{J.~Garc\'ia-Bellido}
\affiliation{Instituto de Fisica Teorica UAM/CSIC, Universidad Autonoma de Madrid, 28049 Madrid, Spain}

\author{G.~Giannini}
\affiliation{Institut de F\'{\i}sica d'Altes Energies (IFAE), The Barcelona Institute of Science and Technology, Campus UAB, 08193 Bellaterra (Barcelona) Spain}

\author{D.~Gruen}
\affiliation{University Observatory, Faculty of Physics, Ludwig-Maximilians-Universit\"at, Scheinerstr. 1, 81679 Munich, Germany}

\author{R.~A.~Gruendl}
\affiliation{Center for Astrophysical Surveys, National Center for Supercomputing Applications, 1205 West Clark St., Urbana, IL 61801, USA}
\affiliation{Department of Astronomy, University of Illinois at Urbana-Champaign, 1002 W. Green Street, Urbana, IL 61801, USA}

\author{G.~Gutierrez}
\affiliation{Fermi National Accelerator Laboratory, P. O. Box 500, Batavia, IL 60510, USA}

\author{S.~R.~Hinton}
\affiliation{School of Mathematics and Physics, University of Queensland,  Brisbane, QLD 4072, Australia}

\author{D.~L.~Hollowood}
\affiliation{Santa Cruz Institute for Particle Physics, Santa Cruz, CA 95064, USA}

\author{D.~J.~James}
\affiliation{Center for Astrophysics $\vert$ Harvard \& Smithsonian, 60 Garden Street, Cambridge, MA 02138, USA}

\author{A.~Kosowsky}
\affiliation{Department of Physics and Astronomy, University of Pittsburgh, Pittsburgh PA 15260 USA}

\author{K.~Kuehn}
\affiliation{Australian Astronomical Optics, Macquarie University, North Ryde, NSW 2113, Australia}
\affiliation{Lowell Observatory, 1400 Mars Hill Rd, Flagstaff, AZ 86001, USA}

\author{M.~Lokken}
\affiliation{David A. Dunlap Department of Astronomy and Astrophysics, University of Toronto, 50 St. George Street, Toronto, Ontario, M5S 3H4 Canada}
\affiliation{Canadian Institute for Theoretical Astrophysics, University of Toronto, 60 St. George St., Toronto, ON M5S 3H4, Canada}
\affiliation{Dunlap Institute of Astronomy \& Astrophysics, 50 St. George St., Toronto, ON M5S 3H4, Canada}

\author{T.~Louis \orcidlink{0000-0002-6849-4217}}
\affiliation{Université Paris-Saclay, CNRS/IN2P3, IJCLab, 91405 Orsay, France}

\author{J.~L.~Marshall}
\affiliation{George P. and Cynthia Woods Mitchell Institute for Fundamental Physics and Astronomy, and Department of Physics and Astronomy, Texas A\&M University, College Station, TX 77843,  USA}

\author{J.~McMahon}
\affiliation{Department of Astronomy and Astrophysics, University of Chicago, Chicago, IL 60637, USA}
\affiliation{Department of Physics, University of Chicago, Chicago, IL 60637, USA}
\affiliation{Kavli Institute for Cosmological Physics, University of Chicago, Chicago, IL 60637, USA}
\affiliation{Enrico Fermi Institute, University of Chicago, Chicago, IL 60637, USA}

\author{J.~Mena-Fern{\'a}ndez}
\affiliation{Centro de Investigaciones Energ\'eticas, Medioambientales y Tecnol\'ogicas (CIEMAT), Madrid, Spain}

\author{F.~Menanteau}
\affiliation{Center for Astrophysical Surveys, National Center for Supercomputing Applications, 1205 West Clark St., Urbana, IL 61801, USA}
\affiliation{Department of Astronomy, University of Illinois at Urbana-Champaign, 1002 W. Green Street, Urbana, IL 61801, USA}

\author{R.~Miquel}
\affiliation{Instituci\'o Catalana de Recerca i Estudis Avan\c{c}ats, E-08010 Barcelona, Spain}
\affiliation{Institut de F\'{\i}sica d'Altes Energies (IFAE), The Barcelona Institute of Science and Technology, Campus UAB, 08193 Bellaterra (Barcelona) Spain}

\author{K.~Moodley \orcidlink{000-0001-6606-7142}}
\affiliation{A. Astrophysics Research Centre, University of KwaZulu-Natal, Westville Campus, Durban 4041, South Africa}
\affiliation{B. School of Mathematics, Statistics \& Computer Science, University of KwaZulu-Natal, Westville Campus, Durban 4041, South Africa}

\author{T.~Mroczkowski \orcidlink{0000-0003-3816-5372}}
\affiliation{European Southern Observatory (ESO), Karl-Schwarzschild-Strasse 2, Garching 85748, Germany}

\author{S.~Naess \orcidlink{0000-0002-4478-7111}}
\affiliation{Institute of Theoretical Astrophysics, University of Oslo. P.O. Box 1029 Blindern, NO-0315 Oslo, Norway}

\author{M.~D.~Niemack \orcidlink{0000-0001-7125-3580}}
\affiliation{Department of Physics, Cornell University, Ithaca, NY 14853, USA}
\affiliation{Department of Astronomy, Cornell University, Ithaca, NY 14853, USA}

\author{R.~L.~C.~Ogando}
\affiliation{Observat\'orio Nacional, Rua Gal. Jos\'e Cristino 77, Rio de Janeiro, RJ - 20921-400, Brazil}

\author{L.~Page}
\affiliation{Joseph Henry Laboratories of Physics, Jadwin Hall, Princeton University, Princeton, NJ, USA 08544}

\author{S.~Pandey}
\affiliation{Department of Physics and Astronomy, University of Pennsylvania, Philadelphia, PA 19104, USA}

\author{A.~Pieres}
\affiliation{Laborat\'orio Interinstitucional de e-Astronomia - LIneA, Rua Gal. Jos\'e Cristino 77, Rio de Janeiro, RJ - 20921-400, Brazil}

\affiliation{Observat\'orio Nacional, Rua Gal. Jos\'e Cristino 77, Rio de Janeiro, RJ - 20921-400, Brazil}

\author{A.~A.~Plazas~Malag\'on}
\affiliation{Department of Astrophysical Sciences, Peyton Hall, Princeton University, Princeton, NJ USA 08544}

\author{M.~Raveri}
\affiliation{Department of Physics, University of Genova and INFN, Via Dodecaneso 33, 16146, Genova, Italy}

\author{M.~Rodriguez-Monroy}
\affiliation{Centro de Investigaciones Energ\'eticas, Medioambientales y Tecnol\'ogicas (CIEMAT), Madrid, Spain}

\author{E.~S.~Rykoff}
\affiliation{Kavli Institute for Particle Astrophysics \& Cosmology, P. O. Box 2450, Stanford University, Stanford, CA 94305, USA}
\affiliation{SLAC National Accelerator Laboratory, Menlo Park, CA 94025, USA}

\author{S.~Samuroff}
\affiliation{Department of Physics, Northeastern University, Boston, MA 02115, USA}

\author{E.~Sanchez}
\affiliation{Centro de Investigaciones Energ\'eticas, Medioambientales y Tecnol\'ogicas (CIEMAT), Madrid, Spain}

\author{E.~Schaan}
\affiliation{Kavli Institute for Particle Astrophysics \& Cosmology, P. O. Box 2450, Stanford University, Stanford, CA 94305, USA}
\affiliation{SLAC National Accelerator Laboratory, Menlo Park, CA 94025, USA}

\author{I.~Sevilla-Noarbe}
\affiliation{Centro de Investigaciones Energ\'eticas, Medioambientales y Tecnol\'ogicas (CIEMAT), Madrid, Spain}

\author{E.~Sheldon}
\affiliation{Brookhaven National Laboratory, Bldg 510, Upton, NY 11973, USA}

\author{C.~Sif\'on \orcidlink{0000-0002-8149-1352}}
\affiliation{Instituto de F\'isica, Pontificia Universidad Cat\'olica de Valpara\'iso, Casilla 4059, Valpara\'iso, Chile}

\author{M.~Smith}
\affiliation{School of Physics and Astronomy, University of Southampton,  Southampton, SO17 1BJ, UK}

\author{M.~Soares-Santos}
\affiliation{Department of Physics, University of Michigan, Ann Arbor, MI 48109, USA}

\author{F.~Sobreira}
\affiliation{Instituto de F\'isica Gleb Wataghin, Universidade Estadual de Campinas, 13083-859, Campinas, SP, Brazil}
\affiliation{Laborat\'orio Interinstitucional de e-Astronomia - LIneA, Rua Gal. Jos\'e Cristino 77, Rio de Janeiro, RJ - 20921-400, Brazil}

\author{E.~Suchyta}
\affiliation{Computer Science and Mathematics Division, Oak Ridge National Laboratory, Oak Ridge, TN 37831}

\author{G.~Tarle}
\affiliation{Department of Physics, University of Michigan, Ann Arbor, MI 48109, USA}

\author{C.~To}
\affiliation{Center for Cosmology and Astro-Particle Physics, The Ohio State University, Columbus, OH 43210, USA}

\author{C.~Vargas \orcidlink{0000-0001-5327-1400}}
\affiliation{ Instituto de Astrof\'isica and Centro de
Astro-Ingenier\'ia, Facultad de F\'isica, Pontificia Universidad
Cat\'olica de Chile, Av. Vicu\~na Mackenna 4860, 7820436 Macul,
Santiago, Chile}

\author{E.~M.~Vavagiakis}
\affiliation{Department of Physics, Cornell University, Ithaca, NY 14853, USA}

\author{N.~Weaverdyck}
\affiliation{Department of Physics, University of Michigan, Ann Arbor, MI 48109, USA}
\affiliation{Lawrence Berkeley National Laboratory, One Cyclotron Road, Berkeley, CA 94720, USA}

\author{J.~Weller}
\affiliation{Max Planck Institute for Extraterrestrial Physics, Giessenbachstrasse, 85748 Garching, Germany}
\affiliation{University Observatory, Faculty of Physics, Ludwig-Maximilians-Universit\"at, Scheinerstr. 1, 81679 Munich, Germany}

\author{P.~Wiseman}
\affiliation{School of Physics and Astronomy, University of Southampton,  Southampton, SO17 1BJ, UK}

\author{B.~Yanny}
\affiliation{Fermi National Accelerator Laboratory, P. O. Box 500, Batavia, IL 60510, USA}

\begin{abstract}

The kinematic and thermal Sunyaev-Zel'dovich (kSZ and tSZ) effects probe the abundance and thermodynamics of ionized gas in galaxies and clusters. We present a new hybrid estimator to measure the kSZ effect by combining cosmic microwave background temperature anisotropy maps with photometric and spectroscopic optical survey data. The method interpolates a velocity reconstruction from a spectroscopic catalog at the positions of objects in a photometric catalog, which makes it possible to leverage the high number density of the photometric catalog and the precision of the spectroscopic survey. Combining this hybrid kSZ estimator with a measurement of the tSZ effect simultaneously constrains the density and temperature of free electrons in the photometrically selected galaxies. Using the 1000 deg$^2$ of overlap between the Atacama Cosmology Telescope (ACT) Data Release 5, the first three years of data from the Dark Energy Survey (DES), and the Baryon Oscillation Spectroscopic Survey (BOSS) Data Release 12, we detect the kSZ signal at 4.8$\sigma$ and reject the null (no-kSZ) hypothesis at 5.1$\sigma$.  This corresponds to 2.0$\sigma$ per 100,000 photometric objects with a velocity field based on a spectroscopic survey with 1/5$^{\textrm{th}}$ the density of the photometric catalog. For comparison, a recent ACT analysis using exclusively spectroscopic data from BOSS measured the kSZ signal at 2.1$\sigma$ per 100,000 objects. Our derived constraints on the thermodynamic properties of the galaxy halos are consistent with previous measurements. With future surveys, such as the Dark Energy Spectroscopic Instrument and the Rubin Observatory Legacy Survey of Space and Time, we expect that this hybrid estimator could result in measurements with significantly better signal-to-noise than those that rely on spectroscopic data alone. 

\end{abstract}
\maketitle

\section{\label{sec:intro}Introduction}

As photons from the cosmic microwave background (CMB) propagate towards Earth, they interact with gas in the intervening space. These interactions leave imprints on the CMB. The two interactions considered in this paper are the thermal and kinematic Sunyaev-Zel'dovich effects (tSZ effect and kSZ effect, respectively) \citep{Sunyaev,Sunyaev:1980nv, hand}. The kSZ effect is caused when low-energy CMB photons Compton scatter off of moving free electrons.  The interaction results in an energy shift that depends on the density and coherent motion of the free electrons, scaling with the projected electron momentum along the line-of-sight. The tSZ effect is also caused by the scattering of CMB photons, but is dependent on the random thermal motion of the electrons as opposed to their bulk motion, thus scaling with the product of the electron temperature and density. Measurements of these effects can teach us about the distribution and thermodynamic properties of the baryons in and between galaxies and clusters, and the growth of cosmological structures.

Previous kSZ measurements \citep{Schaan_2021, Hill_2016, Kusiak_2021} and recent fast radio burst measurements (such as \cite{Macquart_2020}), provide multiple low-z probes of the mean baryon abundance that agree with the CMB and big-bang nucleosynthesis (BBN) cosmic abundance. However, current measurements of the baryon densities within galaxy halos only account for a portion of the overall cosmic abundance of baryons \citep{Fukugita_2004, 2006ApJ...650..573C}. This suggests there must be additional baryons beyond the virial radii of these halos. This uncertainty around the location of the baryons is known as the missing baryon problem. Both SZ effects provide ways to trace baryons in ionized gas at large halo-centric radii and potentially resolve this problem (e.g., \cite{Moodley_2009}, and see \cite{Mroczkowski_2019} for a review).

Unlike X-ray surface brightness, which depends on the square of the gas density, the surface brightness of the kSZ effect has a linear dependence on density. This means that the kSZ effect can be used to trace low-density regions of ionized gas that cannot be easily traced with X-ray observations; consequently kSZ measurements are crucial for understanding galaxy evolution. In particular, they test models of galaxy evolution and feedback that predict different distributions of ionized gas in and around halos \citep{Battaglia_2017, battaglia2019}.

Over the years the kSZ signal has been measured using a variety of techniques. The pairwise-momentum estimator uses CMB temperature maps and galaxy catalogs to measure the net difference in temperature between clusters that are moving towards each other under the influence of gravity \citep{hand}. This estimator can also be applied in fourier-space as demonstrated in \cite{Sugiyama_2018}.
The kSZ signal has also been measured using projected fields \citep{Hill_2016,Ferraro_2016, Kusiak_2021, projected}. This method relies on cross-correlating the square of a CMB temperature map with a projected density map constructed from large-scale structure tracers. Velocity-reconstruction stacking \citep{ Schaan_2016, Tanimura_2021, Tanimura_2022} is another approach that involves stacking CMB temperature maps at the location of galaxies, weighted by the line-of-sight velocities associated with those galaxies. This is the method that we use for this paper.

The first measurement of the kSZ effect was presented in \cite{hand} in 2012 using the pairwise-momentum estimator, cross-correlating Atacama Cosmology Telescope (ACT) maps with spectroscopic galaxy measurements from the SDSS-III Baryon Oscillation Spectroscopic Survey (BOSS) \cite{Eisenstein_2011}. Since then, a number of detections using the pairwise method have been made with ACT and BOSS data, including \cite{Bernardis_2017} and \cite{Calafut_2021}. This method was also applied to DESI imaging surveys and \textit{Planck} in \cite{pairwise_photo} and to the DES redMaPPer cluster catalog and SPT data in \cite{Schiappucci_2023}. The first kSZ measurement with photometric data was presented in \cite{Hill_2016}, which used projected fields to measure the kSZ effect at 3.8-4.5$\sigma$ significance using \textit{Planck}, \emph{WMAP}, and \emph{WISE} data.

Soon after, \cite{Soergel_2016} presented the first pairwise kSZ measurement using photometric data. This was done with maps from the South Pole Telescope and galaxy catalogs from the Dark Energy Survey (DES). Doing so, they were able to detect the kSZ signal at 4.2$\sigma$ and reject the null hypothesis at 2.4$\sigma$.

Recent velocity-reconstruction stacking measurements from ACT and BOSS \cite{Schaan_2021} [hereafter S21] present the highest-significance kSZ measurement to date at 7.9$\sigma$ and rule out the null hypothesis at 6.5$\sigma$. 

These high-significance measurements have helped push the boundaries of what we know about the kSZ effect, by leveraging large amounts of spectroscopic data and high-resolution CMB maps. Except for \cite{Hill_2016,Soergel_2016,Kusiak_2021}, these measurements have been primarily limited to spectroscopic catalogs. This is due to the fact that measuring the kSZ effect generally depends on having very well-constrained 3D locations of the objects in RA, declination, and redshift. For stacking analyses, this is because it is necessary to first estimate the radial velocities of the halos in question. These velocities are typically inferred from the 3D density of galaxies in galaxy catalogs using the continuity equation. Similarly, for pairwise estimators the estimator depends on knowing the distances between pairs, and thus the 3D location of halos. In both cases, these quantities are much more robustly inferred with high-precision spectroscopic redshifts than with photometric redshifts. However, an advantage of photometric surveys is that they typically have significantly higher densities of objects than are found in spectroscopic surveys. It would be advantageous to exploit the much denser photometric catalogs in order to improve our ability to measure and understand the kSZ signal. This idea was explored in \cite{Nguyen_2020}, which used a forward modeling technique to combine spectroscopic galaxy data from SDSS with photometric and spectroscopic catalogs of clusters also from SDSS to measure the kSZ signal at 2$\sigma$.

In this work, we present a new hybrid estimator that involves interpolating the velocity field constructed from spectroscopic galaxy catalogs to infer the velocity field at the locations of galaxies in a photometric survey. This alternative method allows us to leverage the more precise redshift measurements from the spectroscopic catalogs and the higher density of galaxies in the photometric catalogs. For the analysis presented here, the hybrid estimator enables us to use 256023 photometric galaxies, whose velocities we estimate using a significantly smaller sample of 51321 spectroscopic galaxies.   

In addition to studying the kSZ effect, we include a measurement of the tSZ effect \citep{Sunyaev}, which depends on the temperature and density of the electrons. In contrast, the kSZ effect traces the density of the electrons. By combining these two effects, it is possible to tease out an estimate for the electron temperature. We then go on to fit a physically motivated model to the measured SZ signals, which can subsequently be compared to theoretical predictions from hydrodynamical simulations. Throughout this work we compare to the results from S21 and \cite{Amodeo_2021} [hereafter A21] which are companion papers that present a similar, but independent, analysis of the kSZ and tSZ signals associated with the BOSS CMASS objects. Overall, we find that the signals we measure, and the models we fit to these signals, are consistent at the current level of precision. Future, higher-precision data, may be able to probe differences in the gas properties between different galaxy samples. 

This paper is organized as follows. The data that we use for this measurement are presented in \S \ref{sec:data}. The hybrid estimator and our analysis pipeline are presented in \S \ref{sec:analysis}. We present our kSZ measurement in \S \ref{sec: ksz} and our tSZ measurement in \S \ref{sec: tsz}. The electron temperature profile is discussed in \S \ref{sec:electron_temp} and modeling is discussed in \S \ref{sec: model}.  Finally, our discussion is in \S \ref{sec:discussion} and our conclusion is in \S \ref{sec:conclusion}.

\section{\label{sec:data}Data}
\subsection{ACT CMB Maps}
We study the profile of the gas in halos using CMB temperature maps from ACT \citep{Fowler:07, actpol}.  ACT was a 6-meter telescope located in the Atacama Desert in Chile. The telescope first started observing in 2007 and has had a series of upgrades to its camera, the most recent of which is Advanced ACTPol \citep{Henderson2016, 10.1117/12.2233113, Choi_2018, 10.1117/12.2313942}. 

The data from these different instruments were combined as part of ACT's Data Release 5 (DR5) to produce coadded maps that cover nearly half the sky at arcminute resolution \cite{naess2020}. The depth, resolution, and large sky coverage of these maps are well-suited to our kSZ analyses. In particular, we focus on the ACT DR5 2008-18 temperature maps in the f090 and f150 frequency channels. These frequency channels are roughly centered at 98 and 150 GHz but see Figure 2 of \cite{naess2020} for an overview of the passbands. The ACT DR5 maps used here were constructed by coadding multiple seasons worth of ACT data with \emph{Planck} data \citep{Planck-overview:2018}. The resulting maps have beams with full-width half-maxima of 2.1 and 1.3 arcminutes at f090 and f150, respectively. 

For our core analysis, as in S21, we rely on maps containing both data collected in the day and the night (referred to as daytime and nighttime data); however, we also use maps that include just the nighttime data alone for null tests. The inclusion of these extra data sets allows us to make difference maps between the nighttime and day+night maps. These difference maps should contain no signal and can thus be used for a null test. The sky coverage shared by the BOSS, DES, and ACT maps mean that we primarily use data in ACT's Deep 56 (D56) area, which is the area of ACT’s deepest DR5 coverage, with depths of 12–18 and 8–12 $\mu$K arcmin in f090 and f150, respectively. In addition to these maps, we use the inverse variance maps (\textit{ivar} maps), which include an estimate of the non-atmospheric inverse variance in 1/$\mu{\rm K}^2$ per pixel. These allow us to account for variations in survey depth across the ACT sky area when estimating the kSZ signal.

For the tSZ measurement, we use the component-separated Compton-$y$ map presented in \cite{Madhavacheril_2020}, which was constructed from ACT DR4 and \emph{Planck} data. These component-separated maps cover $\sim$ 2,100 deg$^2$ at arcminute resolution and overlap with the sky region used for this analysis. The $y$ maps are constructed using an internal linear combination (ILC) of the ACT maps at f090 and f150, as well as \textit{Planck} maps at 30, 44, 70, 100, 143, 217, 353, and 545 GHz, which fill in missing low-$\ell$ modes that ACT does not measure well and also covers frequency channels that ACT does not observe in. We use the Compton-$y$ maps with a fiducial cosmic infrared background (CIB) spectral energy distribution (SED) deprojected to reduce contamination from extragalactic dust. This choice was motivated by null tests in S21 that demonstrated their sensitivity to CIB contamination in their tSZ measurements (see fig 22 of S21). The CIB deprojection is described in detail in \cite{Madhavacheril_2020} but to summarize, a modified blackbody spectrum is deprojected from the ILC map via a constrained ILC technique \citep{Remazeilles_2011}, assuming an effective temperature of $T_{CIB} = $ 24 K and spectral index $\beta =$ 1.2.  The resulting map has a Gaussian beam with FWHM = 2.4\textquotesingle .

\subsection{BOSS CMASS Catalog}
As in S21, we use the CMASS (``constant stellar mass'') sample from BOSS DR12 \citep{BOSS, DR12}; this is a galaxy catalog from the Sloan Digital Sky Survey (SDSS). The catalog is selected such that the stellar mass limit is constant over redshift, and it includes galaxies with spectroscopic redshift measurements ranging from 0.4 to 0.7 \cite{BOSS}. 

\begin{figure}   
    \includegraphics[width = \linewidth]{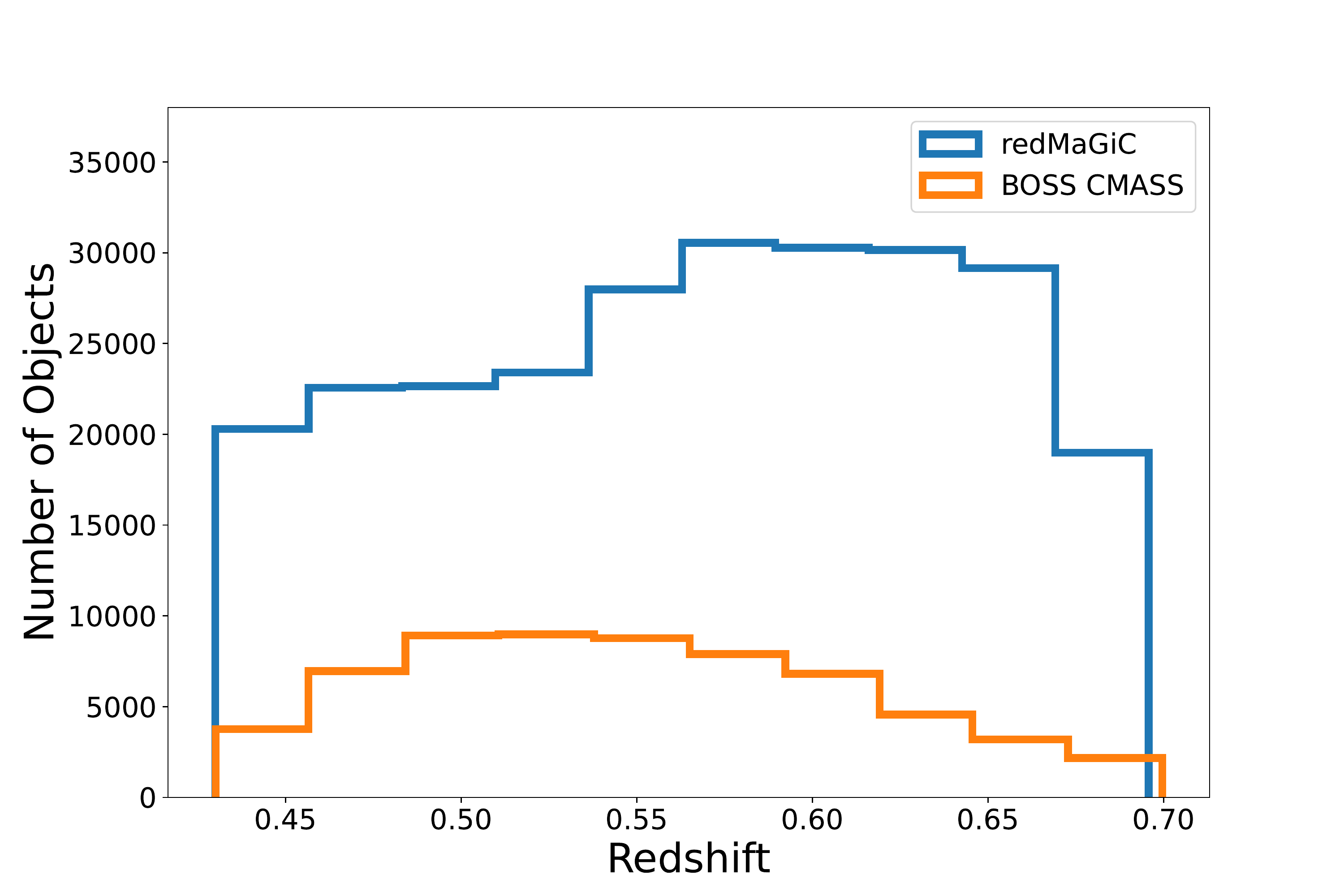}
    \caption{Redshift distribution of the trimmed DES redMaGiC and BOSS CMASS catalogs. This includes only the 51321 BOSS and 256023 redMaGiC objects that are used in our final analysis.\label{fig: redshift_dist}}
\end{figure}

For this analysis, we limit the sample to those objects that overlap with the DES redMaGiC and ACT DR5 sky area. This results in a sample of 51321 objects.  The redshift distribution is presented in Figure \ref{fig: redshift_dist} and Figure \ref{fig: footprint} presents the sky region, which has $\sim$ 6000 deg$^2$ overlapping between BOSS and ACT but just 1000 deg$^2$ between all three surveys.

We use the reconstructed radial velocity measurements used in S21 and described in \cite{Vargas_Maga_a_2016}. The measurements use the 3D number density of galaxies, smoothed with a fixed smoothing scale of 5$h^{-1}$\,Mpc, to infer their line-of-sight velocity. This is done by solving the continuity equation (\cite{Padmanabhan_2012}, S21):
\begin{equation} \label{eq: cont}
\bm{\nabla}\cdot \bm{v}
+ f  \bm{\nabla}\cdot \left[ \left( \bm{v}\cdot \bm{\hat{n}} \right) \bm{\hat{n}} \right]
= - \frac{a H f \delta_g}{b} \ .
\end{equation}
Here the galaxy over-density is $\delta_g$, the logarithmic linear growth rate is $f$, $\bm{v}$ is the velocity vector, $\bm{\hat{n}}$ is the line-of-sight direction, $H$ is the Hubble parameter, and $b$ is the linear galaxy bias. This method works by placing the galaxies in a 3D box that is significantly larger than the sky area covered by the sample to map out the density field. It then assumes linear theory such that the velocity field is a gradient of the scalar field. Initially, this assumption is valid because the vector component of the initial velocity field decays with the inverse of the scale factor. At late times this assumption remains reasonable because, although nonlinear evolution under gravity sources a vector component, this component is small on the large scales that we probe, and so we can neglect it here.

We interpolate the reconstructed velocity field from \cite{Vargas_Maga_a_2016} to estimate the velocity field at the locations traced by our photometric sample, described in \S \ref{sec: redmagic}; this process is described in \S \ref{sec:analysis_estimator}.

\subsection{DES redMaGiC Catalog}\label{sec: redmagic}
DES \citep{DES_main} is a photometric survey in the $grizY$ bands that covers $\sim$ 5000 deg$^2$ using the Dark Energy Camera \citep{2015AJ....150..150F}, which is mounted on the Blanco 4-meter Telescope at
the Cerro Tololo Inter-American Observatory (CTIO) in the Chilean Andes. From the full survey, DES selects red galaxies with their red-sequence
Matched-filter Galaxy Catalog algorithm (redMaGiC, \cite{2016MNRAS.461.1431R}). This results in a catalog of objects with optimal photometric redshift data that is well-suited for SZ measurements. 

\begin{figure}
    \centering
    \includegraphics[width = \linewidth]{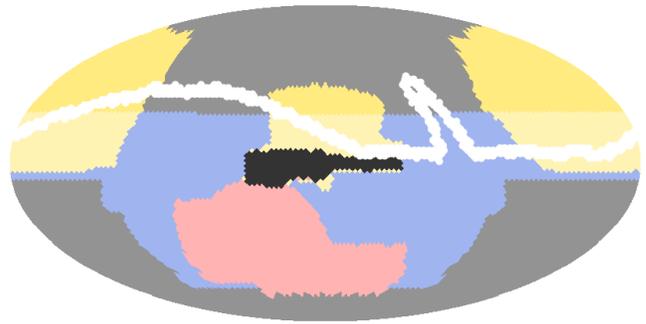}
    \caption{Overlap of the ACT DR5 maps (blue), the DES redMaGiC catalog (red), and the BOSS CMASS catalog (yellow), plotted in equatorial coordinates. The overlapping region for all three surveys is shown in black and constitutes 1000 deg$^2$. We also show the outline of the upcoming LSST survey (below the white line), which covers a similar region as DELVE . For future studies we note that the overlap of ACT and BOSS constitutes 5737 deg$^2$ and ACT, BOSS, and LSST will cover 3773 deg$^2$.}
    \label{fig: footprint}
\end{figure}

Here, we use the DES-Year-3 (DES-Y3) high-density redMaGiC galaxy catalog, which includes the first three years of DES data. This data has well-calibrated photo-$z$ measurements with a mean redshift error of approximately 0.02 \cite{Sevilla_Noarbe_2021}. After limiting the catalog to the region that overlaps with the BOSS CMASS footprint and the ACT DR5 map area we are left with 256,023 objects (approximately five times more than in the CMASS sample) as shown in Figure \ref{fig: redshift_dist}. This sample has a mean redshift of 0.57 and the overlapping area between the three surveys is approximately 1000 deg$^2$, as shown in Figure \ref{fig: footprint}. In Figure \ref{fig: footprint} we also give an outline of the sky area covered by other surveys including LSST and DELVE (DECam Local Volume Exploration Survey) \citep{Drlica_Wagner_2021}, which could be used to expand upon this work in the future. Finally we estimate an average halo mass for this sample of $M_H \approx 10^{13.4\pm 0.1} \, M_{\odot}$ based on \cite{zacharegkas2021dark}, which models the weak gravitational lensing signal of halos hosting redMaGiC galaxies with a halo occupation distribution framework (see Section \ref{sec:gnfw} for more details).

\begin{table}
\centering
\def\arraystretch{1.5}
\setlength\tabcolsep{5pt}
\begin{tabular}{|c|c|c|c|}
\hline
Data Set & \shortstack{Total number \\ of galaxies }&\addstackgap{\shortstack{Galaxies in\\ overlapping \\ sky area}} & \shortstack{Total\\sky area} \\
\hline
\addstackgap{\shortstack{ACT DR5 \\frequency maps }}&-& -& 18,000  \\
\addstackgap{\shortstack{ACT DR4 \\$y$ maps}} &-&- & 2,100 \\
BOSS CMASS & 777,202&51,321& 10,000    \\
DES redMaGiC &2,766,815& 256,023& 5,000 \\
\hline
\end{tabular}
\caption{The number of galaxies and the sky area available for each survey. All sky areas are listed for the unmasked maps and are in deg$^2$. The total overlapping sky area is 1000 deg$^2$. \label{table: data}}
\end{table}

\section{\label{sec:analysis}Analysis}
\begin{figure}
    \centering
    \includegraphics[width = \linewidth, trim={9cm 2cm 8cm 3cm},clip ]{"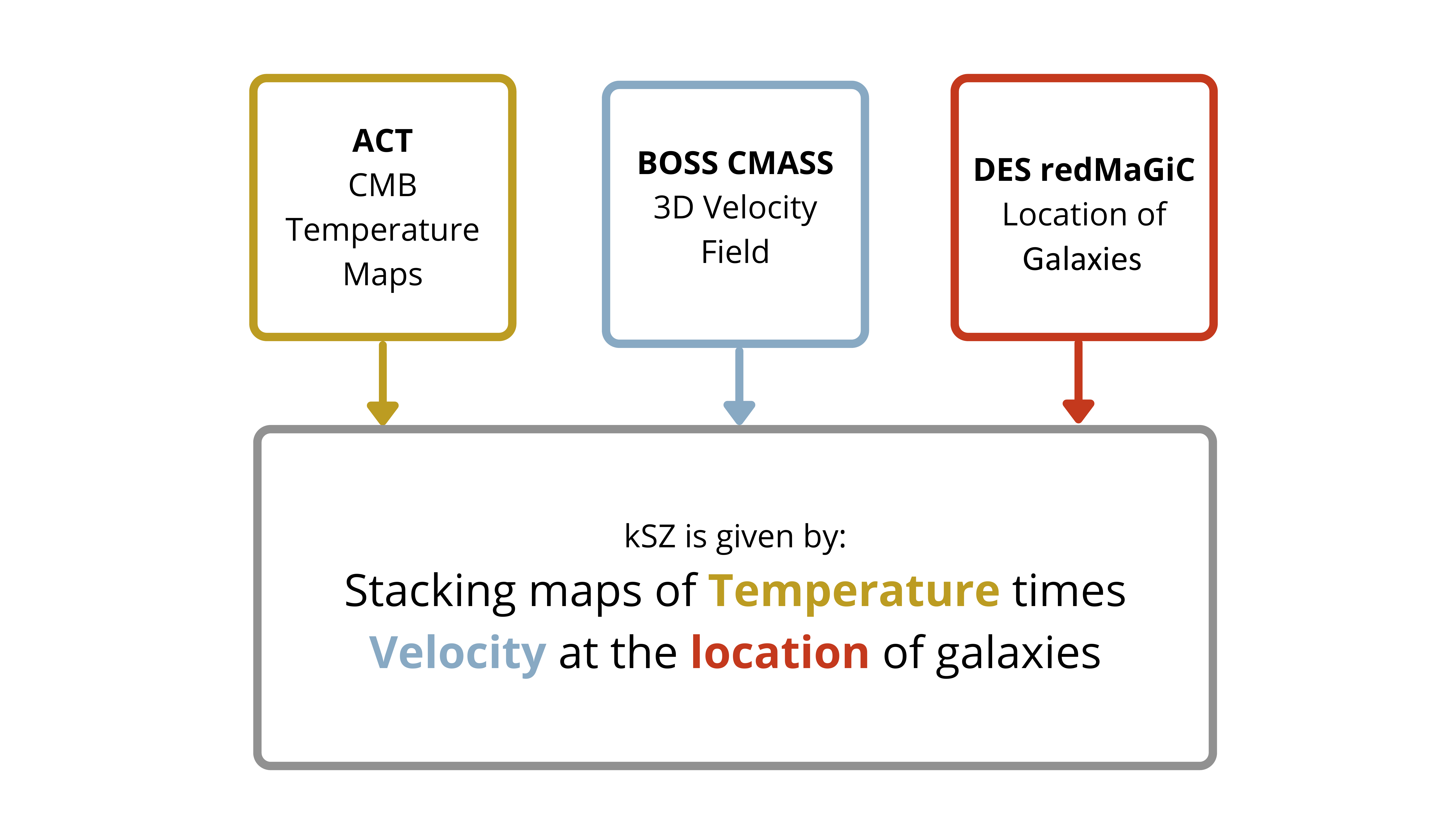"}
    \caption{The hybrid estimator presented in this work relies on combining three data sets in order to measure the kSZ effect. Here we outline the process that goes into this measurement.}
    \label{fig:process}
\end{figure}

We measure the kSZ effect using spectroscopic data to estimate the 3D velocity field, photometric data to trace the location of galaxies, and CMB data to measure the kSZ signal. This process is summarized in Figure~\ref{fig:process} and described in detail in the subsections that follow.

\subsection{The Hybrid Velocity Estimator}\label{sec:analysis_estimator}
One of the challenges associated with kSZ measurements is obtaining accurate velocity estimates for the objects in question. Typical methods for velocity estimation use the 3D density of objects to infer their velocities \citep{Vargas_Maga_a_2016}. These methods are typically limited to spectroscopic data because they rely heavily on the accuracy of the redshift measurements. However, in many cases photometric catalogs contain significantly more objects, so it would be beneficial to make use of these larger catalogs. 

We use the velocity reconstruction from the BOSS catalog to estimate the 3D velocity field. We then evaluate the velocity of this field at the 3D location of DES redMaGiC galaxies. As such, the photometric redshifts of the redMaGiC objects are only used to determine the three-dimensional position of the objects, which means that the relatively large photometric redshift errors (in comparison to those from spectroscopy) only lead to small errors in the redMaGiC velocity estimates. For comparison, directly estimating the velocity field from photometry could result in much larger errors since it requires taking the derivative of noisy data\footnote{Refs.~\cite{Keisler_2013} and \cite{Soergel_2016} both studied the effect of using photometric data in kSZ measurements using the pairwise estimator and found that the photo-$z$ errors play a significant role in suppressing the kSZ amplitude. We note that while the hybrid estimator presented here provides one way of combining spectroscopic and photometric data, a better alternative could be to model the velocity field using both data sets and one overall forward modeling pipeline. Such an approach could build upon the work presented in \cite{arxiv.2210.15649} and \cite{Prideaux_Ghee_2022}.}. We find that, at the mean redshift of our sample, the average photo-$z$ error corresponds to an error in the 3D localization of less than 50 Mpc. In comparison, the velocity fields are coherent on scales of $\sim$100 Mpc, a factor of two larger. This makes it possible to use our hybrid estimator, which relies on the long coherence lengths of the velocity field and well-calibrated photo-$z$, to estimate radial velocity values for the redMaGiC galaxies from the BOSS velocity field.

We implement this method using \texttt{nbodykit} \citep{Hand_2018} to convert the coordinates of each object in RA, DEC, and redshift to Cartesian positions. To do so, we assume a fiducial $\Lambda$CDM cosmology with parameters from the \textit{Planck} 2015 CMB analysis~\cite{2016}\footnote{This is the same cosmology used in the BOSS velocity reconstruction.}. Next, we project the BOSS objects onto a 3D grid of voxels that measure 10 Mpc in each dimension. We then interpolate over this regular grid to reconstruct the velocities of the redMaGiC objects. 

This interpolation process will inherently lead to some bias in the resulting velocities. Traditionally, these biases are analyzed using simulations such as in \cite{Vargas_Maga_a_2016}. However, in order to perform that type of test, it would be necessary to have a simulation that included both a CMASS-like and redMaGiC-like sample in the same simulation. In addition to the overlapping sample, the simulation would need to have a similar number density of objects, given that these reconstructions rely heavily on understanding the 3D density field. Currently, such a simulation is not available, meaning a direct analysis of the bias is difficult, and we instead perform partial tests of the reconstruction where possible.

One test we include is to examine the efficacy of the interpolation method we use in our reconstruction. We do this using exclusively the BOSS CMASS objects, for which we already have velocity estimates. We take the sample of CMASS objects and split it into a test sample, which contains 10\% of the data and base sample, which contains the remaining 90\% of the data. We then interpolate over the base sample to reconstruct the velocities of the test sample. We refer to the original reconstructed velocities from \cite{Vargas_Maga_a_2016} as the ``Original Velocities" and the velocities calculated through our interpolation method as the ``Interpolated Velocities". 

\begin{figure}
    \centering
    \includegraphics[width=\linewidth]{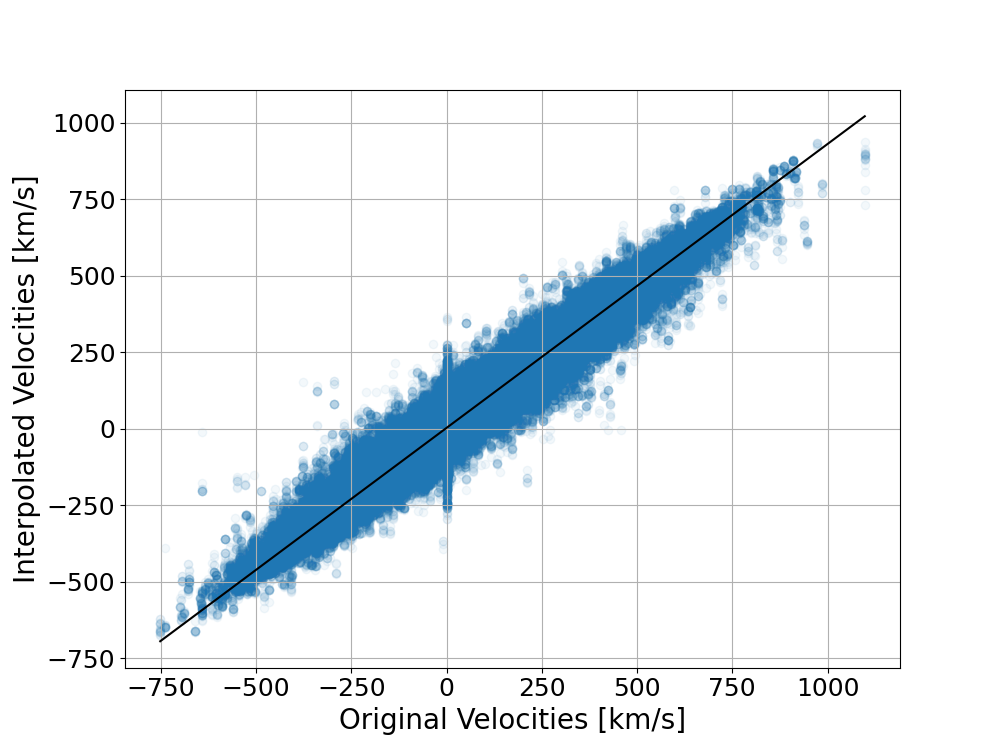}
    \caption{We verify the efficacy of our interpolation method by testing our ability to recover the velocities for the objects in the BOSS CMASS catalog. We find that the correlation coefficient for this test is 0.95 and the bias (defined as bias = interpolated velocities/original velocities - 1) is -0.07.}
    \label{fig: boss_reconstruction}
\end{figure}

Once we have the interpolated velocities, we can measure the correlation between the original velocities and the interpolated velocities. This process is then repeated 100 times, randomly splitting the sample each time, and the results are shown in Figure~\ref{fig: boss_reconstruction}. We find that the process produces a sample with a correlation coefficient of 0.95. We also look at the bias from this test, defined as bias = interpolated velocities/original velocities - 1, and find that it is -0.07. We test propagating this bias through to our kSZ measurement below and find that the effect is an order of magnitude smaller than the errors on our measurement.

\subsection{Aperture Photometry and the SZ signals}\label{sec:analysis_AP}
We measure the stacked radial kSZ and tSZ profiles of the redMaGiC galaxies in the ACT maps by using aperture photometry (AP) filters following the procedure in S21. To do so, we apply AP filters of varying radii to each object, which yield the integrated radial profiles of the signals. 

To apply a filter, we integrate over a disk of angular radius $\Theta$ centered on the halo in question. We then subtract off an annulus with an outer radius of $\sqrt{2}\Theta$ that surrounds the disk. The resulting radial profile is given by:

\begin{equation}\label{eq:AP}
    T_{AP}(\Theta) = \int_0^\Theta \delta T(\theta) d\theta - \int_\Theta^{\sqrt{2}\Theta} \delta T(\theta) d\theta \,,
\end{equation}
where $\delta T$ is the temperature map. AP filters are particularly useful for kSZ measurements because they reduce the noise from CMB fluctuations, which cannot be separated from kSZ fluctuations using multifrequency information, as both signals have the same blackbody spectrum.  The AP filter mitigates noise from primary CMB fluctuations with wavelengths longer than the size of the filter, as these modes are effectively removed when we subtract off the annulus. 

We apply these filters by first selecting  20$\times$20 arcminute cut-outs of the CMB maps centered on each of the DES galaxies. For the tSZ measurements, we extract similar cut-outs from Compton-$y$ maps with the CIB deprojected. For the kSZ measurements, we  use ACT DR5 CMB maps. In both cases, we take care to reproject our CMB maps such that the projection is centered on each galaxy. This mitigates distortions caused by taking cut-outs of the maps from different declinations in the Plate Carre\'{e}-projected ACT maps. We then apply the AP filters to each of the cut-outs as well as to the inverse-variance maps. These profiles are then used to calculate the kSZ signal, discussed in Section \ref{sec: ksz}, and the tSZ signal, discussed in Section \ref{sec: tsz}.

\subsection{Covariance Matrix}

For both the tSZ and kSZ effects, we calculate covariance matrices by bootstrap resampling our data as was done in S21. To do so we generate catalogs by selecting galaxies, with repetition, from our sample such that our bootstrap catalog has the same number of objects as our original one. Next we calculate the associated SZ signals by repeating the stacking and weighting process outlined in Sections \ref{sec: ksz} and \ref{sec: tsz} and then repeat the process 10,000 times\footnote{We found that the $\chi^2$ measurements associated with the covariance matrices stabilized at 8,000 bootstrap iterations, thus we allowed for 10,000 iterations to ensure these numbers had converged.}. We then calculate the covariance of the bootstrapped samples in order to get the relevant covariance matrices. For the kSZ signal, we calculate the covariance matrix of the combined f090 and f150 profiles, which we can then use to calculate the combined signal-to-noise of our results.

\begin{figure}
    \centering
    \includegraphics[width=\linewidth, trim={0 0 2.5cm 0},clip ]{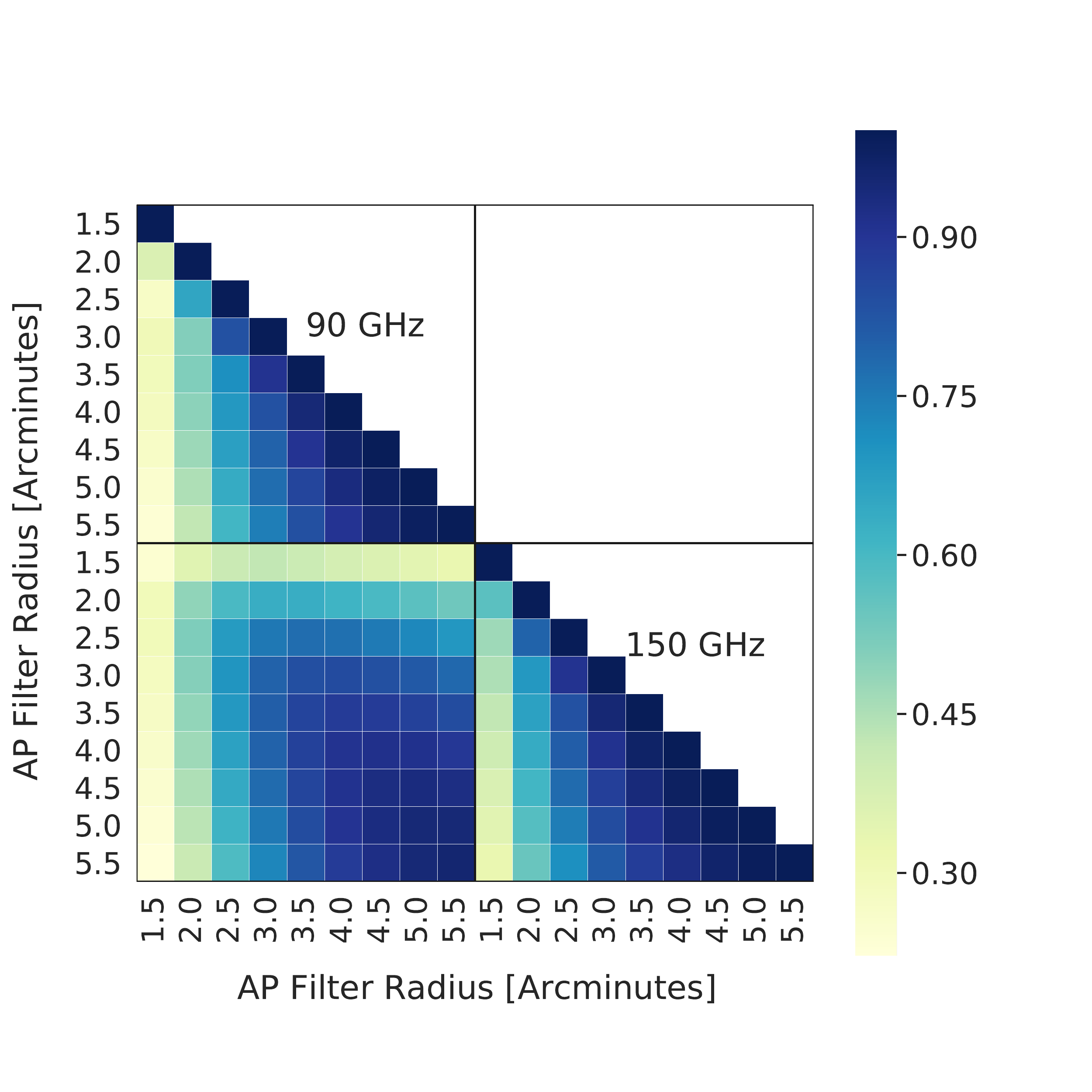}
    \caption{The f090 and f150 correlation matrix calculated by bootstrapping the kSZ data. This is equivalent to Figure 7 in S21 and shows similarly high correlations at large radii.}
    \label{fig: cor}
\end{figure}

Figure \ref{fig: cor} shows the correlation matrix calculated from this covariance. The bins at larger radii are highly correlated due to the aperture photometry filtering. In Appendix \ref{appendix: cor} we test whether these high correlations affect the expected number of degrees of freedom and conclude that the expected number of degrees of freedom is consistent with the number of spatial degrees of freedom.

We calculate the signal-to-noise ratio (SNR) by first calculating the $\chi^2$ as:
\begin{equation*}
    \chi^2 = (d - m)^tC^{-1}(d - m) \,,
\end{equation*}
where $C$ is the covariance matrix, $d$ represents the data, and $m$ represents the given model. The model is either null, when considering rejection of the null hypothesis, or our best-fit model. To convert our $\chi^2_{\rm null}$ into SNR, we first calculate the probability-to-exceed (PTE), which we then express in terms of Gaussian standard deviations,

\begin{equation}
    (1 - \textrm{PTE}) = \textrm{erf}(\textrm{SNR}/\sqrt{2}) \,,
\end{equation}
where erf refers to the error function.

For the best-fit models, we calculate the SNR using the same approach as in S21, which is:
\begin{equation}\label{eq:snr}
    \textrm{SNR} \equiv  \sqrt{\Delta \chi^2_{\textrm{null} - \textrm{best-fit}}} =  \sqrt{\chi^2_{\textrm{null}}- \chi^2_{\textrm{best-fit}}} \,.
\end{equation}
This quantity corresponds to the significance of the preference for the best-fit model over the null hypothesis.

\section{The \lowercase{k}SZ Signal} \label{sec: ksz}
Since the kSZ effect is dependent on velocity, the $T_{AP}(\Theta)$ needs to be weighted by velocity in order to avoid canceling the signal. The oscillating signs of the line-of-sight velocity also means that by weighting our stack by the velocity we make the estimator highly robust against foregrounds, which are velocity independent. We use the following estimator, which was proposed in S21:
\begin{equation}\label{eq:ksz}
    T_{\rm{kSZ}}(\Theta) = \frac{v^{\rm{rms}}_{\rm{rec}}}{c}\frac{\sum_i T_{AP_i}(\Theta)(v_{{\rm rec},i}/c)w_i}{\sum_i (v_{{\rm rec}, i}/c)^2w_i}\, ,
\end{equation}
where $v_{\rm{rec}, \textit{i}}$ is the mean-subtracted line-of-sight velocity for object $i$ reconstructed with the hybrid technique and $v^{\rm{rms}}_{\rm{rec}}$ is the RMS of this velocity. We also define inverse-variance weights using $w_i = 1/(\sigma_{z,i}^2\sigma_{m,i}^2$). Here we account for the noise in the CMB maps with $\sigma_{m,i}$, which is calculated by applying the AP filters to the \textit{ivar} maps, such that each individual halo has its own value for $\sigma_{m,i}$. This allows us to account for the large distribution in noise across the ACT DR5 maps. The $\sigma_{z,i}$ is the redshift error in comoving coordinates associated with each object, and it allows us to preferentially weight objects that have low redshift errors when stacking. We choose to combine the two weights in order to optimize for both the CMB-map-based noise and the noise associated with the location of each redMaGiC object.

This equation would ordinarily include a scaling factor of $\frac{1}{r_{v}}$, which is the correlation between the true and reconstructed velocities and is used to account for bias caused by the velocity reconstruction method. While we know the bias for the reconstruction of the BOSS catalog, as discussed in Section \ref{sec:data} we do not have measurements for the full bias of the velocity reconstruction of the photometric DES data. For this reason, we include this as a free parameter (defined as the scale factor $S$) in our model instead of trying to estimate it and include it in our kSZ calculation. Future analyses using this method could look at measuring this number using simulations. We note that characterizing this factor more precisely is necessary to accurately determine the amplitude of our profile; however, it should not affect the profile shape or the SNR of our measurement, but instead just suppresses the amplitude of the signal. 

Excepting the inclusion of $\sigma_{z,i}$ used to account for the redshift errors and the exclusion of the velocity bias, Equation \ref{eq:ksz} is equivalent to the one used to calculate the kSZ signal in S21.

\begin{figure}
    \centering
    \includegraphics[width=\linewidth]{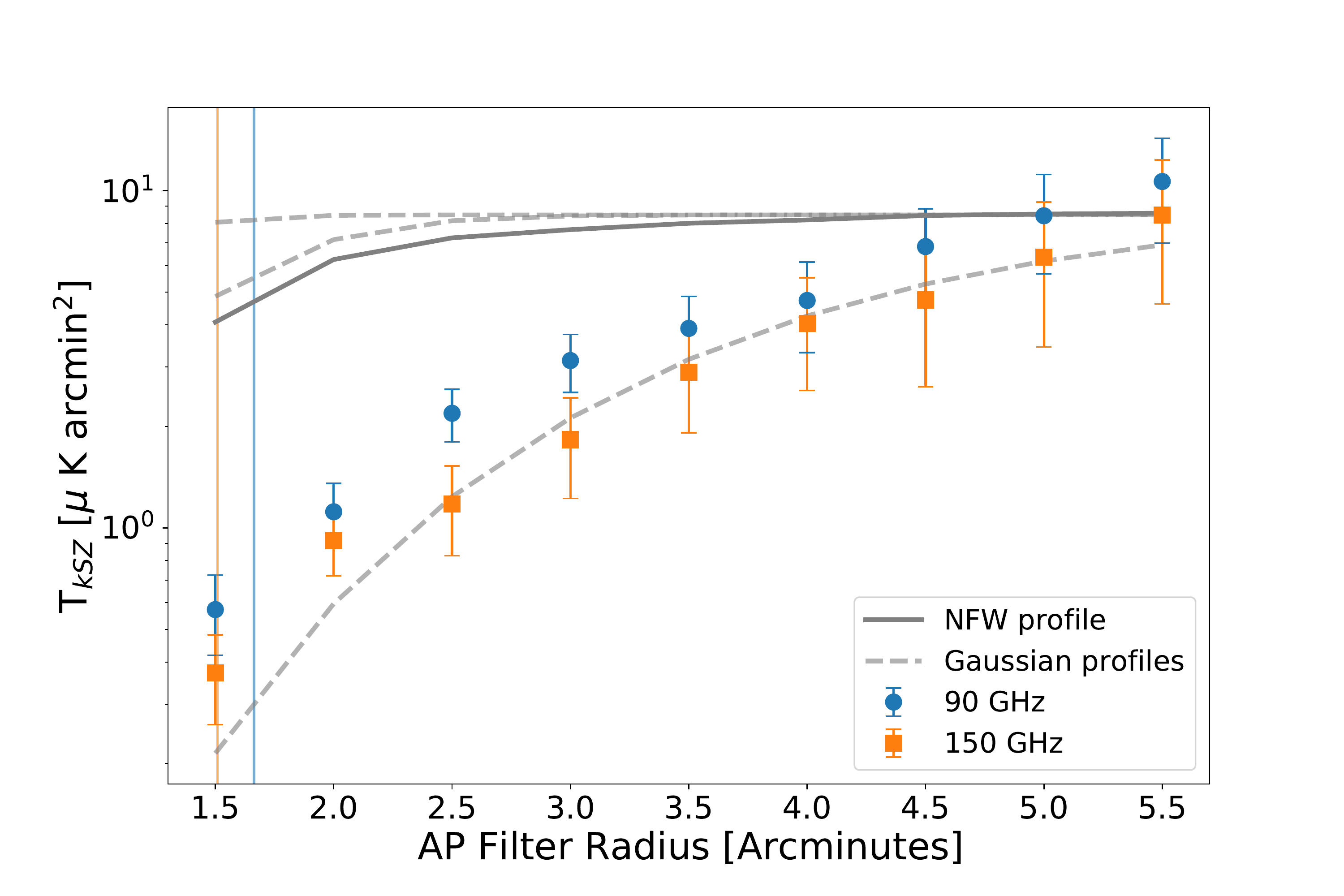}
    \caption{Stacked profile of the kSZ signal of DES redMaGiC galaxies in units of $\mu$K arcmin$^2$. The scaling on the y-axis includes bias from the velocity reconstruction that is later included as a free parameter in the model shown in Figures \ref{fig:individualfreq} and \ref{fig:joint_ksz}, but not accounted for here. The solid gray line is an NFW profile, which assumes the gas follows the dark matter. The two dashed gray lines near the NFW profile are Gaussian profiles with FWHM of 1.3\textquotesingle\  and 2.1\textquotesingle, respectively. The gray line that most closely follows the data represents a Gaussian profile that is much more extended with FWHM = 6\textquotesingle, this means that the data prefer a gas profile that is more extended than the NFW. These are normalized to the signal in the largest aperture for f150. We also plot the virial radius (1.4\textquotesingle\ at $z = 0.57$), calculated based on the mean mass of the sample, added in quadrature with the beam, as a vertical blue line for the f090 beam and as an orange line for the f150 beam. 
    }
    \label{fig:kSZ}
\end{figure}

By applying our pipeline to the f090 and f150 maps separately, we obtain the results shown in Figure \ref{fig:kSZ}. 
These profiles allow us to rule out the null hypothesis at 5.1$\sigma$ for the combined data, and 4.9$\sigma$ and 3.7$\sigma$ for the f090 and f150 bands, respectively. For both frequencies, this data is shown with the beam unconvolved. The presence of the beams should lead to a small difference between the profiles at f090 and f150, despite the kSZ signal having the same SED at both. We would expect the larger f090 beam to result in a slight suppression of the f090 compared to the f150 data. Instead, what we see is a slight positive fluctuation for the f090 data. This fluctuation is explored later in this paper.

In addition to ruling out the null hypothesis, we follow S21 in comparing the data to a simple Navarro-Frenk-White profile (NFW)~\citep{1996ApJ...462..563N}, which assumes that the gas follows the dark matter. To do so we use the mass-concentration relation laid out in \cite{2008MNRAS.390L..64D} to simulate a 3D NFW profile at the mean halo mass of the sample, which we estimate as $10^{13.4} \, M_\odot$ as discussed in Section \ref{sec: redmagic}. This profile is then converted to a number density of free electrons; to do so, we assume the gas is fully ionized with primordial helium abundance and cosmological baryon abundance. The 3D profile is truncated at one virial radius and then  projected to a 2D plane at the mean redshift of the sample. We then convolve it with the ACT f150 beam and apply the AP filtering. We note that, as shown in Figure \ref{fig:kSZ}, the data disfavor this NFW model $(\sqrt{\chi^2_{\textrm{NFW}} - \chi^2_{\textrm{best-fit}}} = 58)$; a similar result was shown in S21.

We also show that the data indicate the presence of an extended gas profile compared to the NFW. This extended profile can also be seen by comparing the data to the three Gaussian profiles shown in gray dashed lines in Figure \ref{fig:kSZ}. The profiles, from highest to lowest, represent Gaussians with FWHM of 1.3\textquotesingle, 2.1\textquotesingle,  and 6\textquotesingle\ that have been passed through the AP filters and then normalized to the last radial bin of the f150 profile. The first two profiles approximate the ACT beam sizes.

It is evident that the data prefer a gas profile that is more extended than the NFW, which is consistent with predictions that the kSZ traces gas in the warm hot intergalactic medium (WHIM) in an extended halo around galaxies \cite{Dave_2001_whim}. It is worth noting that some of the signal we measure at large radii could also be due to contributions from neighboring halos or nearby satellites. We consider these neighboring halos in more detail when we model the kSZ signal in \S \ref{sec: model}, and discuss comparisons with the results of S21 and A21.

\section{The \lowercase{t}SZ Signal}\label{sec: tsz}
For the tSZ measurements, we use the same sky area and DES objects used for the kSZ measurement in order to maintain consistency. The measurement is simpler since it is not necessary to take into account radial velocities; instead we just account for the noise as given by:
\begin{equation}\label{eq:tsz}
    T_{\rm{tSZ}}(\Theta) = \frac{\sum_i T_{AP_i}(\Theta)/\sigma_{m,i}^2}{\sum_i 1/\sigma_{m,i}^2}\,.
\end{equation}
For the sake of interpretation we can convert the above $T_{\rm{tSZ}}(\Theta)$ from $y$ units to CMB temperature units by multiplying by the following:
\begin{equation}\label{eq:convert_y_tsz}
     f_{\rm{tSZ}}(\nu) = x \textrm{coth}(x/2) - 4 \,,
\end{equation}
where $x = \frac{h\nu}{k_BT_{\rm{CMB}}}$.
Using the above relation at f150, a value $y =1$ corresponds to $- 2.59 \times 10^6 \, \mu$K.

We note that the Compton-$y$ maps do not include inverse variance maps, meaning that in this case $\sigma_{m,i}$ is estimated by calculating the variance from the mean of the map for each AP filter radius and for each object $i$.

\begin{figure}
    \centering
    \includegraphics[width = \linewidth]{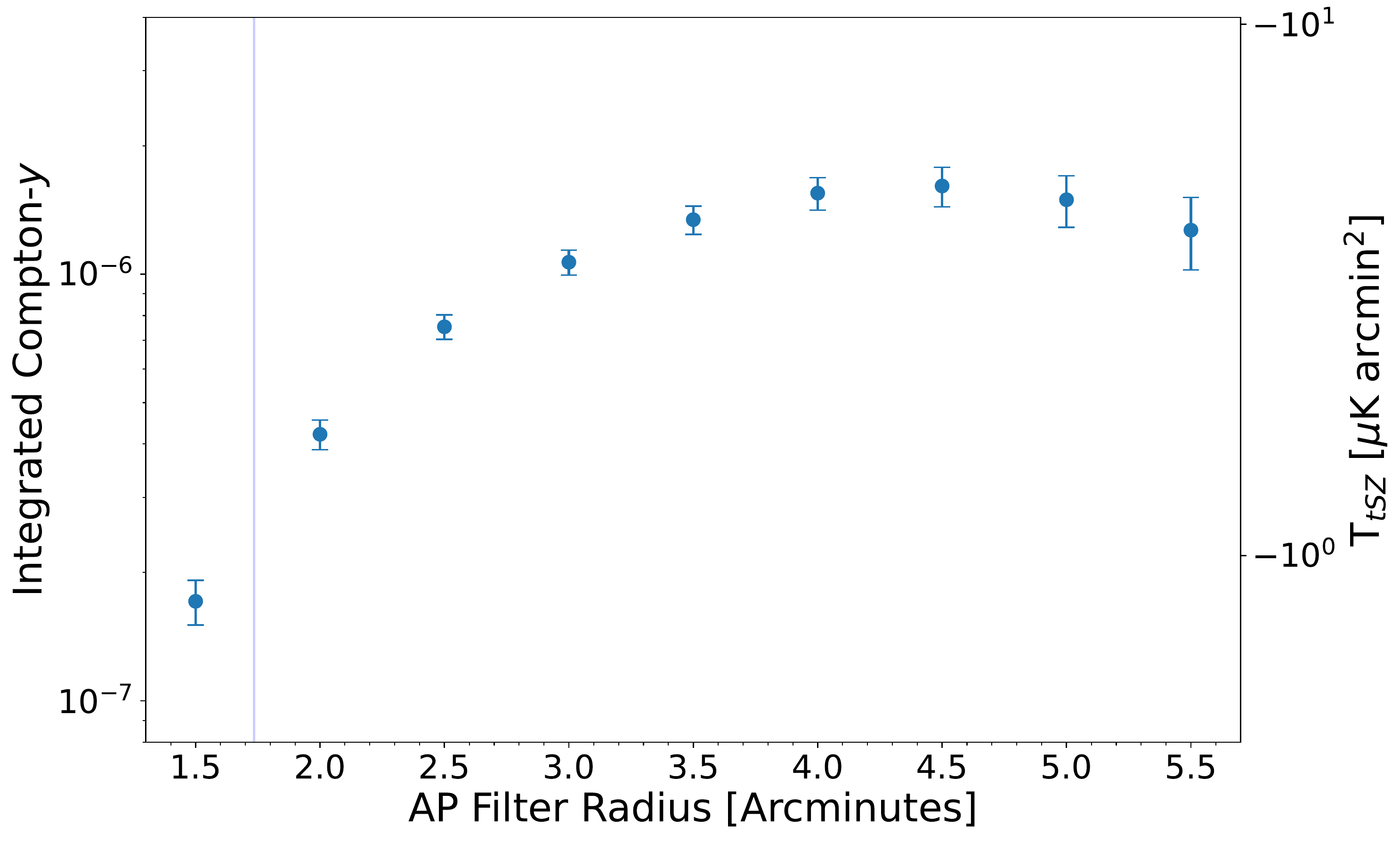}
    \caption{The tSZ profile of DES redMaGiC galaxies, measured using Compton-$y$ maps with a fiducial CIB SED deprojected. We plot the profile in dimensionless Compton-$y$ units on the left-hand side and on the right-hand side we give the signal in temperature units at f150. We convert from $y$ to $T_{\rm{tSZ}}$ using Equation \ref{eq:convert_y_tsz}. We also plot the virial radius, added in quadrature with the beam, as a vertical blue line.}
    \label{fig:tsz}
\end{figure}

By applying the tSZ pipeline to the Compton-$y$ map, we measure the tSZ signal as shown in Figure \ref{fig:tsz}. The tSZ signal is generally an order of magnitude stronger than the kSZ signal, and so is much easier to detect. For our data, we are able to rule out the null hypothesis at 15$\sigma$. 

\section{\label{sec: null} Null Tests}
\begin{figure}
    \centering
    \includegraphics[trim = 1cm 0cm 1cm 1cm, clip,width = \linewidth]{"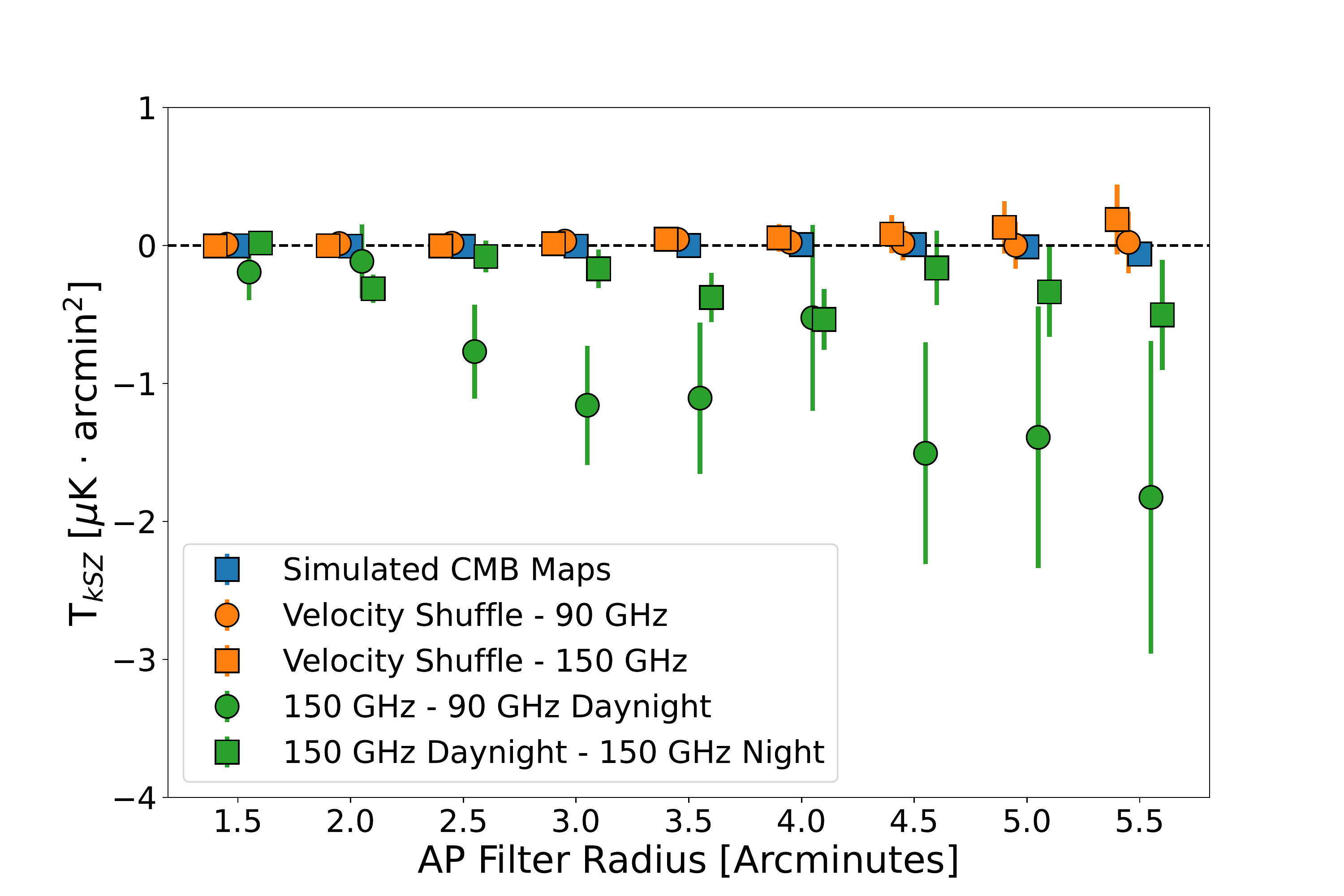"}
    \caption{This figure shows the three types of kSZ null tests we perform. Type 1 (shown in blue): Gaussian map tests performed by simulating CMB maps, passing those maps through the analysis pipeline. Type 2 (shown in orange): shuffled velocities associated with each object. Type 3 (shown in green): we pass difference maps, produced by taking the difference between the day + night maps and the night only maps as well as the difference between f090 and f150 maps, through the pipeline. Table \ref{table: pte} presents the PTE for each null test.}
    \label{fig: null test}
\end{figure}

We test the results of our hybrid estimator pipeline by performing a suite of null tests in both the f090 and f150 channels. The results of these tests are shown in Figure \ref{fig: null test}.

For our first null test (Type 1 in Figure \ref{fig: null test}) we produce 100 simulated CMB-only maps based on a fiducial power spectrum using HEALPIX's \texttt{synfast} function \citep{Gorski_2005}. We then pass these maps through our pipeline (i.e., apply the AP filters, weight with the velocities, and calculate the signal). 

The second null test (Type 2) is based on shuffling the velocities associated with each object. We perform this test on our data by repeatedly scrambling the velocities associated with each object and then calculating the kSZ profile with these scrambled velocities. We repeat this test 250 times for each frequency channel and average over the results. 

Finally, we perform a series of null tests based on taking difference maps, referred to as Type 3. ACT DR5 includes maps using both day time and nighttime data, as well as nighttime data only in both f090 and f150. We take the nighttime and the day+night versions of the maps and reconvolve them to the same beam in order to account for the broader day+night beam. We then take the difference of the two maps. We also take the difference between the f090 and f150 maps, again reconvolved to the same beam. We pass these two difference maps through our pipeline and measure the resulting signal.

For the simulated CMB maps and the velocity shuffle null test we base the error bars on the covariance between multiple iterations of each test. For the difference maps, we instead rely on bootstrapping to measure the covariance matrix, similar to what is done for the fiducial analysis. For all of the tests, we compute the PTE (shown in Table \ref{table: pte}), convert that to equivalent Gaussian standard deviations, and then accept null tests if they pass at 3$\sigma$ (PTE $>$ 0.00275). We find that using this benchmark all of our null tests pass. 

We note that the f150 - f090 test passes our null test, but has a slightly low PTE of 0.09 and visually seems to deviate the furthest from zero in Figure \ref{fig: null test}. This seems to suggest that we may not be fully nulling out the signal when we difference the two maps. This small difference could potentially be attributed to small-scale issues in the DR5 maps, possibly related to contamination from the source subtraction algorithm. Alternatively, it could indicate an issue with the beam profiles we used, or possibly some type of contamination from a frequency-dependent signal. Interestingly, the same null test performed in S21 showed a similar trend. We measure our lowest PTE of 0.03 for the daynight-night null test and we see a small indication of a residual for this test in Figure \ref{fig: null test}. This could indicate a slight calibration error, which would be due to the fact that the DR5 maps are not fully calibrated as discussed in \cite{naess2020}. Additionally, as stated in \cite{naess2020}, these DR5 maps include data products that were not fully characterized and as such gain/beam errors on the order of a few percent are expected in the night maps and slightly larger errors on the order of 10\% may be present for the day+night maps. 

\begin{table}
\def\arraystretch{1.5}
\setlength\tabcolsep{5pt}
\begin{tabular}{|c|c|}
\hline
Null Test & PTE\\
\hline
Difference map: f150 - f090 daynight & 0.09\\
Difference map: f150 daynight - night & 0.03\\
Velocity shuffle - f090 & 0.95 \\
Velocity shuffle - f150 &  0.86 \\
Simulated CMB maps & 0.97\\
\hline
\end{tabular}
\caption{The PTE values for the null tests shown in Figure \ref{fig: null test}. Both sets of difference maps have slightly low PTEs. These low values could be indicative of some inconsistencies at the map or beam level between these datasets, and are discussed in more detail in Section \ref{sec: null} and in Section \ref{sec:combfreq}.  \label{table: pte} }
\end{table}

\section{Electron Temperature Profile}\label{sec:electron_temp}

The kSZ signal is dependent upon $n_e$, the electron number density. In comparison, the tSZ signal depends on both $n_e$ and $T_e$, the electron temperature. By leveraging this relationship, we can estimate the electron temperature as a function of radius for these filters. 

The electron temperature is effectively given by the ratio between the two profiles. Or more specifically, we can use:
\begin{equation}
    T_e = \left(\frac{m_e c^2}{k_B}\right)\left(\frac{y_{AP}}{\tau_{AP}}\right)\,.
\end{equation}
In this equation $y_{AP}$ represents the signal from our AP filtered Compton-$y$ maps and $\tau_{AP} =(c/v_{\rm{rms}}^{\rm{true}})T_{\rm{kSZ}}/T_{\rm{CMB}}$, where $v_{\rm{rms}} = 314$ km/s is the RMS of the peculiar velocities along the line-of-sight for the mean redshift of our sample ($z = 0.57$), assuming a linear relation between velocity and density (as in A21). Because the measurements of $y_{AP}$ and $\tau_{AP}$ are done using different maps, the beam differences between the maps needs to be accounted for. To do so, we reconvolve the f150 temperature map to the wider beam of the $y$ map before applying the filters and calculating $\tau_{AP}$. The resulting profile has the same AP dependencies as our kSZ and tSZ profiles, meaning that it should not be interpreted as a radial differential profile, but instead as an integrated mean profile.

Because the amplitude of the kSZ profile depends on the correlation coefficient of the velocity reconstruction, this measurement of the temperature depends on that coefficient. For that reason, we assume the profile has some overall scaling factor ($S$, see \S \ref{sec: ksz} for details) that would affect the amplitude but not the shape of the profile. If this factor was less than 1, which is what we expect, then this would result in a reduction in the amplitude of the temperature profile. 

\begin{figure}
    \centering
    \includegraphics[width=\linewidth]{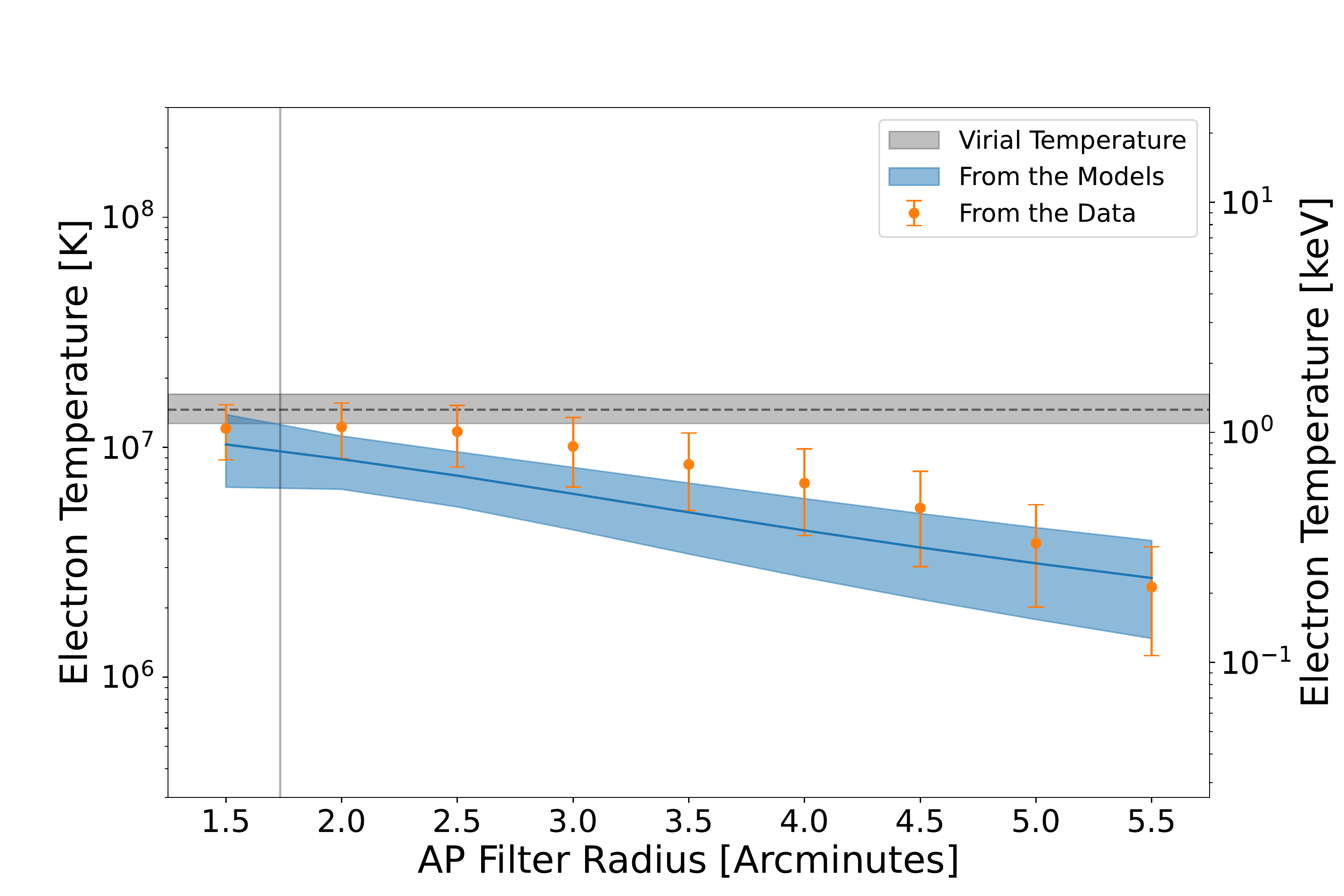}
    \caption{The electron temperature profile given by the ratio of the kSZ and tSZ measurements. We also show the temperature derived from the tSZ and kSZ models, which is based on the spread of the posteriors for those models (and described in Section\ref{sec:elec_model}). For comparison, we plot the virial temperature as a horizontal dashed line with a 1$\sigma$ error band as estimated from the error on the virial mass. The vertical line gives the virial radius, added in quadrature with the beam.}
    \label{fig:etemp}
\end{figure}

The temperature profile, shown in Figure \ref{fig:etemp}, decreases with radius. This shape indicates that while the virial temperature may be a reasonable estimate for the central temperature, there is evidence that the temperature decreases towards the edge of the halos  beyond the virial radius. A similar profile was seen in A21 and S21 when they measured the electron temperature profile for their sample; however their measurement showed evidence for a more rapid drop in temperature.

\section{\label{sec: model} Modeling the SZ Signals}

\subsection{The GNFW Profile \label{sec:gnfw}}
We model the SZ signals presented in \S \ref{sec: ksz} and \S \ref{sec: tsz} using a three-dimensional generalized Navarro-Frenk-White (GNFW) model following A21 and using the convention in \cite{gnfw} \footnote{This is implemented using the Mop-c GT code (\url{https://github.com/samodeo/Mop-c-GT}) presented in A21.}. This model describes the electron number density ($n_e$), which is related to the kSZ signal by the following equation:
\begin{equation}\label{eq: dens}
\frac{\Delta T_{\rm{kSZ}}}{T_{\rm{CMB}}} = \frac{\sigma_T}{c} \int_{los} e^{-\tau} n_e v_p dl \ ,
\end{equation}
where $los$ refers to the line-of-sight, $v_p$ is the line-of-sight peculiar velocity, and $\tau$ is the optical depth to Thomson scattering.

The model also describes the electron pressure profile, which is related to the tSZ signal by:
\begin{equation}
    \frac{\Delta T_{\rm{tSZ}}}{T_{\rm{CMB}}} = f_{\rm{tSZ}}(\nu)y \ ,
\end{equation}
where $f_{\rm{tSZ}}(\nu)$ describes the frequency dependence of the tSZ signal and $y$ is given by:
\begin{equation}
    y(\theta) = \frac{\sigma_T}{m_ec^2}\int_{los} P_e (\sqrt{l^2 + d_A(z)^2|\theta|^2}) dl\ .
\end{equation}
Here $m_e$ is the electron mass and $\sigma_T$ is the Thomson cross-section.

For both profiles, we use the same formalism given in A21. Thus, the GNFW density profile is given by:
\begin{align}
\begin{split}
    \rho(x) &= \rho_0(x/x_{c,k})^{\gamma_k}[1 + (x/x_{c,k})^{\alpha_k} ]^{-\frac{\beta_k-\gamma_k}{\alpha_k}} \,. \\
    \rho_{gas}(x) &= \rho_{cr}(z)f_b\rho(x) \ .
\end{split}
\end{align}
Here $\rho_0$ is the central density, $\rho_{cr}(z)$ is the critical density of the Universe at redshift
z and $f_b$ is the baryon fraction. We define $x = r/R_{200}$ where  $M_{200}$ is the mass within $R_{200}$, within which the halo density is 200 times $\rho_{cr}(z)$. $x_{c,k}$ is the core scale; $\alpha_k$ is the slope at $x \sim$ 1 while $\beta_k$ and $\gamma_k$ give the slopes at $x>>1$ and $ x<< 1$, respectively. 

The thermal pressure profile is given by:
\begin{align}
    \begin{split}
        P(x) &= P_0(x/x_{c,t})^{\gamma_t}[1 + (x/x_{c,t})^{\alpha_t}]^{-\beta_t} \,.\\
        P_{th}(x) &= P(x)P_{200}\,.\\
        P_{200} &= GM_{200}\frac{200 \rho_{cr}(z) f_b}{2R_{200}}.
    \end{split}
\end{align}

For these relations $P_{th}$ is the thermal pressure, the core scale is given by $x_{c,t}$ and, just as for the density, $\alpha_t, \beta_t$ and $\gamma_t$ give the slopes at $x \sim 1$, $x>>1$ and $ x<< 1$, respectively. 

We also include a two-halo term, which is used to account for excess signal from correlated neighboring halos. We implement the method used in \cite{two-halo-10.1093/mnras/stw3311} and A21. This method constructs the two-halo term by considering an average neighboring halo and computing its contribution to the halo-density cross-power spectrum\footnote{See  Appendix A and the related Figure 7 of A21 for details.}. This results in a two-halo term that is dependent on the mean redshift and halo mass of our sample, which is then included in the model with an overall amplitude factor, A$_{2H}$ for the density profile and A$_{P2H}$ for the pressure profile. We assume an average halo mass of $M_H = 10^{13.4} \, M_{\odot}$. This is based on the results of \cite{zacharegkas2021dark}, which quotes an average halo mass per redshift bin. We take those averages and weight them by the number of objects in each redshift bin from our sample to estimate the overall average halo mass. Ideally this two-halo term would be estimated based on the mass and redshift of each individual object in the catalog instead of the mean redshift and mass; however, for this dataset we have a limited understanding of the mass. This limitation means that there could be a certain amount of noise, both in the actual mass measurement and in the redshift binning used to find the mean mass, that could affect our estimate of the two-halo term and affect our estimate for the amplitude of the term later on. For a full discussion of the two-halo term, see \cite{two-halo-10.1093/mnras/stw3311} and A21. 

We then combine the GNFW model with the two-halo term to arrive at our model for the density profile,
\begin{equation}\label{eq: 2halo}
    \rho(x) = \rho_{1H}(x) + \textrm{A}_{2H}\rho_{2H}(x) \,,
\end{equation}
and for the pressure profile,
\begin{equation}
    P(x) = P_{1H}(x) + \textrm{A}_{P2H}P_{2H}(x) \,.
\end{equation}
These models generate three-dimensional profiles that we project into a two-dimensional plane, convolve with the ACT beam profiles, and then apply AP filters, following the procedure used to process the data. When fitting the models, we use Markov Chain Monte Carlo calculations (MCMC) \citep{doi:10.1063/1.1699114} and the \texttt{emcee} algorithm \citep{Foreman_Mackey_2013}. We assume convergence has been reached when the chain length reaches at least 20 times the auto-correlation time, as recommended in \cite{Foreman_Mackey_2013}. 

In the following sections, we present a few different approaches to fitting this model and in all cases we measure the SNR of the model by comparing the preference for the model to the preference for a null signal, as written in equation \ref{eq:snr}.

\subsection{kSZ Individual Frequency Fits}

We begin by fitting the f090 and f150 data independently of each other. In order to do it we follow the procedure in A21, for the kSZ profiles, and we fix $\gamma_k = -0.2$ and $\alpha_k = 1$. From there we assume the same uniform priors used in A21 for $x_{c,k}$ and $\beta$ but slightly wider uniform priors for $\rho_0$ and $\textrm{A}_{2H}$: 1 $<$ log$_{10}\rho_0 <$ 10, 0.1 $< x_{c,k}$ $<$ 1.0, 1 $< \beta_k <10$, and 0 $< \textrm{A}_{2H} <$ 15. The priors for $x_{c,k}$ and $\beta$ are set by physical constraints; $x_{c,k}$ is a ratio of the core and halo radii and as such must be a fraction and $\beta$ is the outer slope which must be larger than the intermediate slope set by $\alpha_k$ and small enough to be physically reasonable. For log$_{10}\rho_0$ we set the lower bound to avoid negative densities and then allow for a high enough upper bound for the parameter to be constrained. The two-halo amplitude is set such that it cannot be negative and the upper limit is set higher than that of S21 to accommodate the data's preference for a larger two-halo term.

To account for the potential bias in the amplitude of the kSZ profile discussed in Section \ref{sec: ksz} we include a scale parameter, $S$, with uniform prior of 0 $<S\leq$ 1.

\begin{figure}
\centering
\subcaptionbox{The f090-only GNFW profile fit for the kSZ profile.\label{fig:90_fit}}
{\includegraphics[width = \linewidth]{"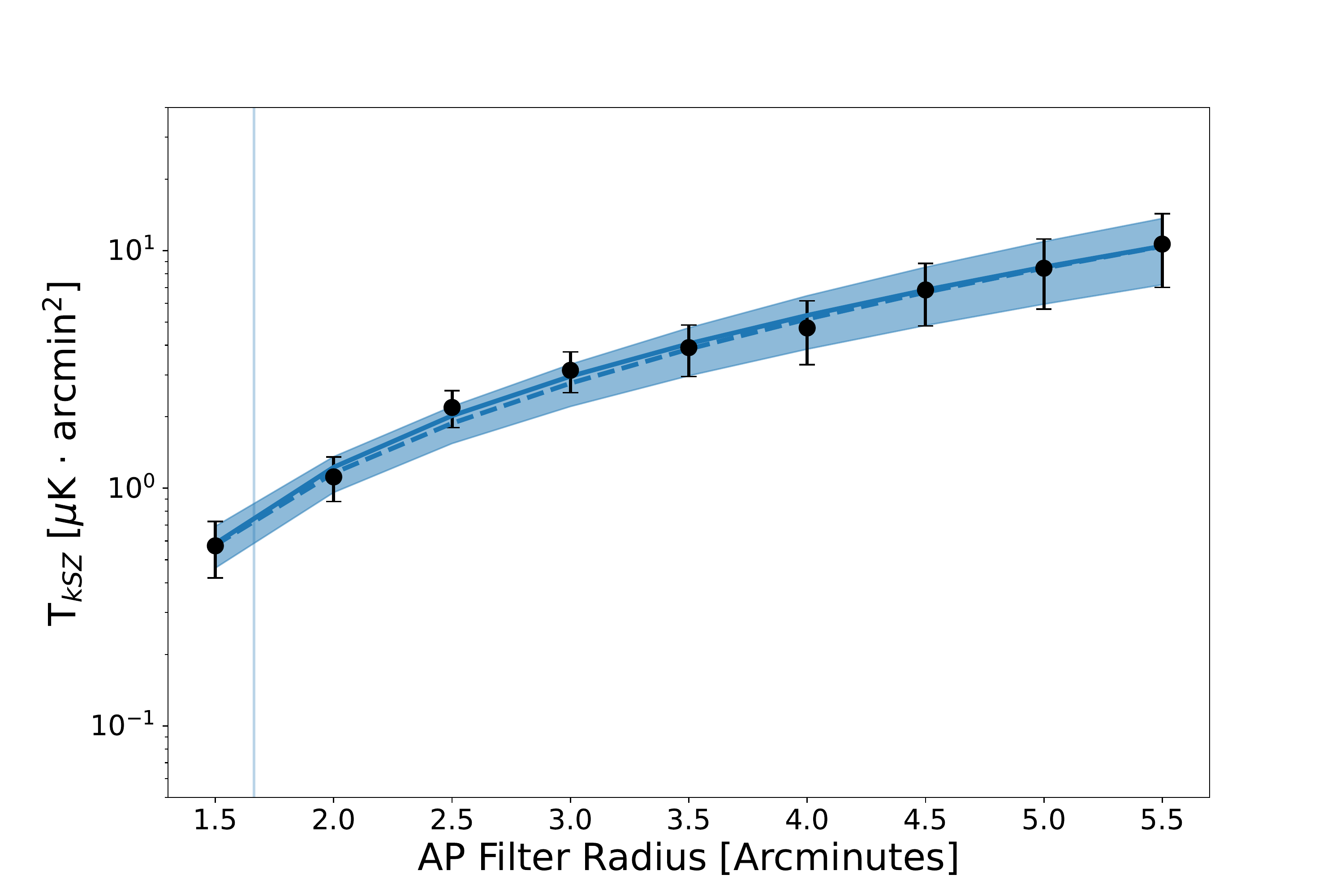"}}
\subcaptionbox{The f150-only GNFW profile fit for the kSZ profile.\label{fig:150_fit}}
{\includegraphics[width = \linewidth]{"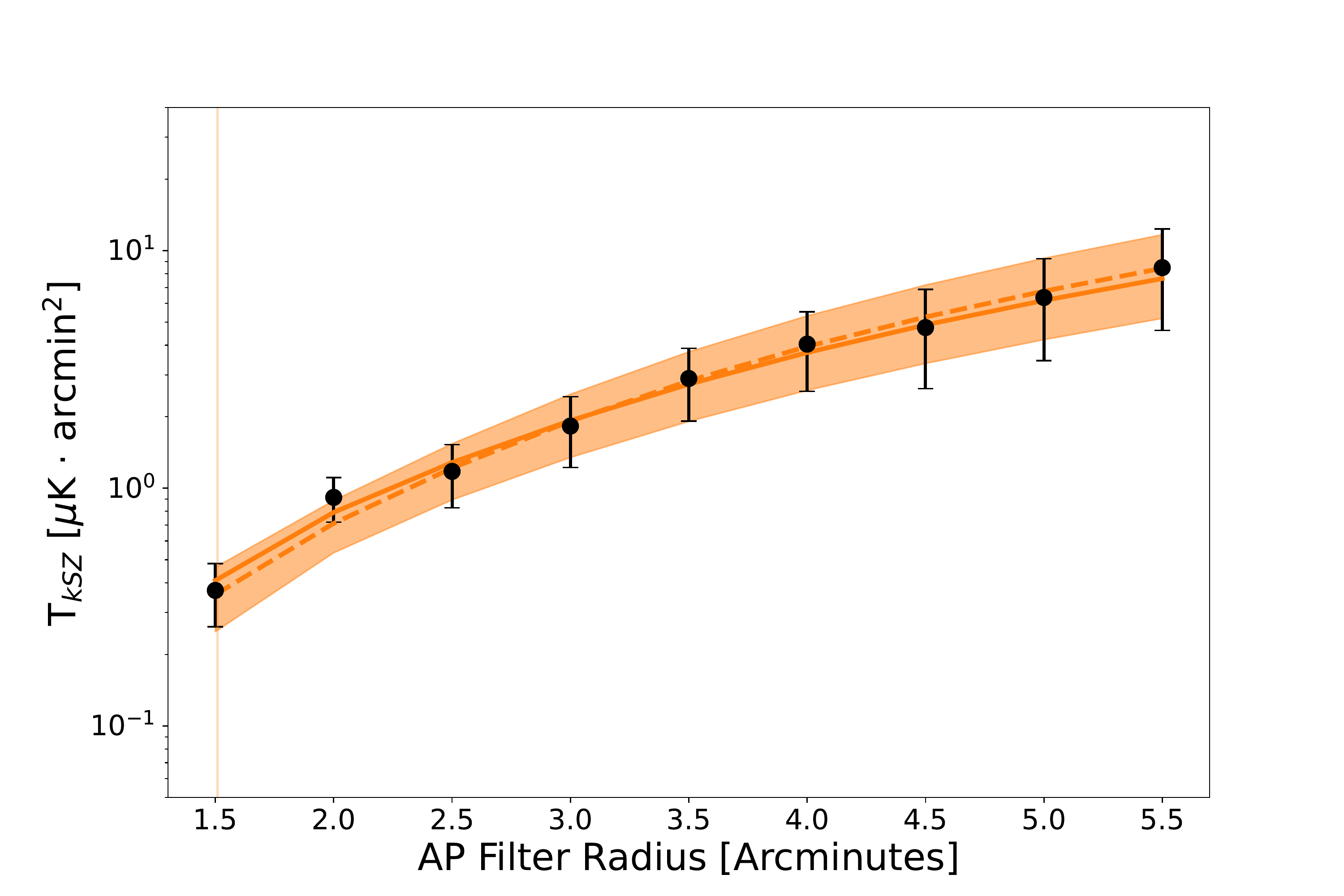"}}
\caption{For both figures, the band gives the 1$\sigma$ spread of the posterior, while the dashed line gives the median and the solid gives the maximum-likelihood solution. We also show the virial radius, added in quadrature with the beam, as a vertical blue line. At f090 and f150 respectively, the maximum-likelihood solution is preferred over the null hypothesis at 6.2$\sigma$ and 4.6$\sigma$  as defined by Equation \ref{eq:snr}\label{fig:individualfreq}.}
\end{figure}

\begin{figure}
    \includegraphics[width = \linewidth]{"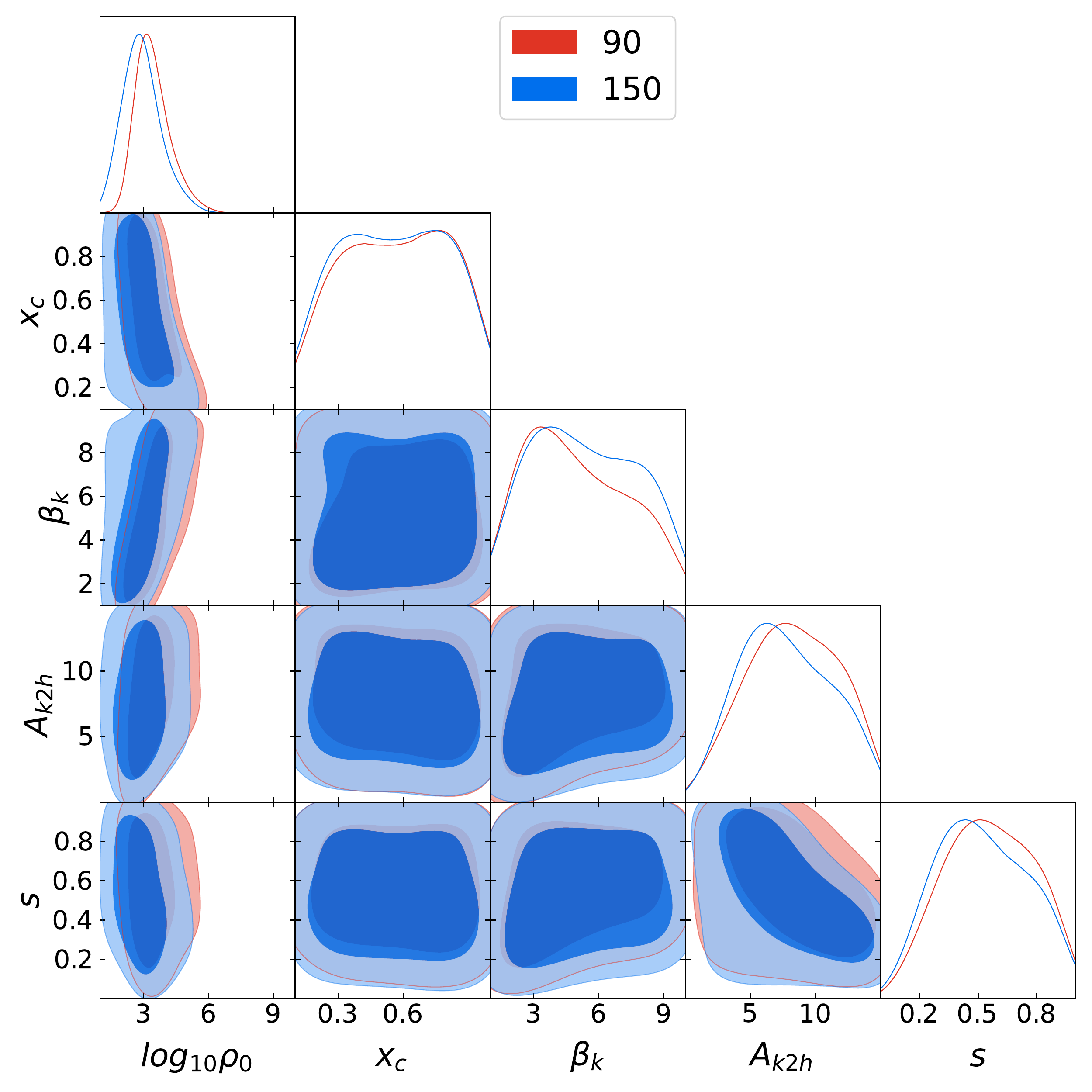"}
    \caption{The independent f090 (red) and f150 (blue) posteriors for the kSZ models in Figure \ref{fig:individualfreq}. We note that, although the parameters are not well-constrained overall, the parameters for the two frequencies are consistent.}
    \label{fig:90_150_like}
\end{figure}

In Figures \ref{fig:90_fit} and \ref{fig:150_fit} we present the fit to the f090 and f150 data, respectively. The best-fit models are preferred over the no-signal hypothesis at 6.2$\sigma$ and 4.6$\sigma$, respectively. For these fits, we show the posteriors in Figure \ref{fig:90_150_like}. We find that our fits are mainly prior dominated, particularly for the core radius $x_c$ and the outer slope $\beta_k$.

The posteriors of the individual fits show good agreement between the f090 and f150 data, and all the parameters are in statistical agreement between the two fits.

\subsection{kSZ Combined Frequency Fits \label{sec:combfreq}}

Given that we sample the same population of objects at both f090 and f150, it is also possible to fit the model for the combined data. The kSZ signal is frequency-independent in CMB thermodynamic temperature units, such that the expected signal should be consistent between the two frequency channels. The only difference between the two channels is the beam profiles. For this reason, when simultaneously fitting the model to both frequency channels, we fit one density profile to both channels convolved with the f090 and f150 beams.

When fitting both the f090 and f150 data simultaneously, we take into account the off-diagonal blocks of the covariance matrix, which account for the covariance between the f090 and f150 data. 

\begin{table}
\centering
\def\arraystretch{1.5}
\setlength\tabcolsep{5pt}
\begin{tabular}{|c|c|c|c|c|}
\hline
GNFW  & &\multicolumn{3}{c|}{ }\\
Density & Priors &\multicolumn{3}{c|}{Constraints}\\\cline{3-5}
Parameters & & f090 & f150 & Combined\\
\hline
$\log_{10}\rho_0$ &[1, 10]    & $3\pm1$             & $3\pm 1$            & 2.1\\
$x_c$             &[0.1, 1.0] & $0.6\pm 0.3$        & $0.6 \pm 0.3$       & 1.0\\
$\beta_k$         &[1, 10]    & $5^{+3}_{-2} $      & $5 \pm 3$           & 3.9\\
$\textrm{A}_{2H}$          &[0, 15]    & $8\pm 4$            & $7^{+5}_{-3}$       & 4.1\\
$S$               &[0, 1]     & $0.6^{+0.3}_{-0.2}$ & $0.5^{+0.3}_{-0.2}$ & 0.7 $\pm$ 0.1\\
\hline
\end{tabular}
\caption{The marginalized constraints for the individual and combined frequency kSZ fits. For the combined fit we fixed the four shape parameters to the best-fit values for the minimum $\chi^2$ solution found using \texttt{scipy optimize}. \label{table: gnfw - ksz} }
\end{table}

\begin{figure}
    \includegraphics[width = \linewidth, trim = 1cm 0cm 1cm 1cm, clip]{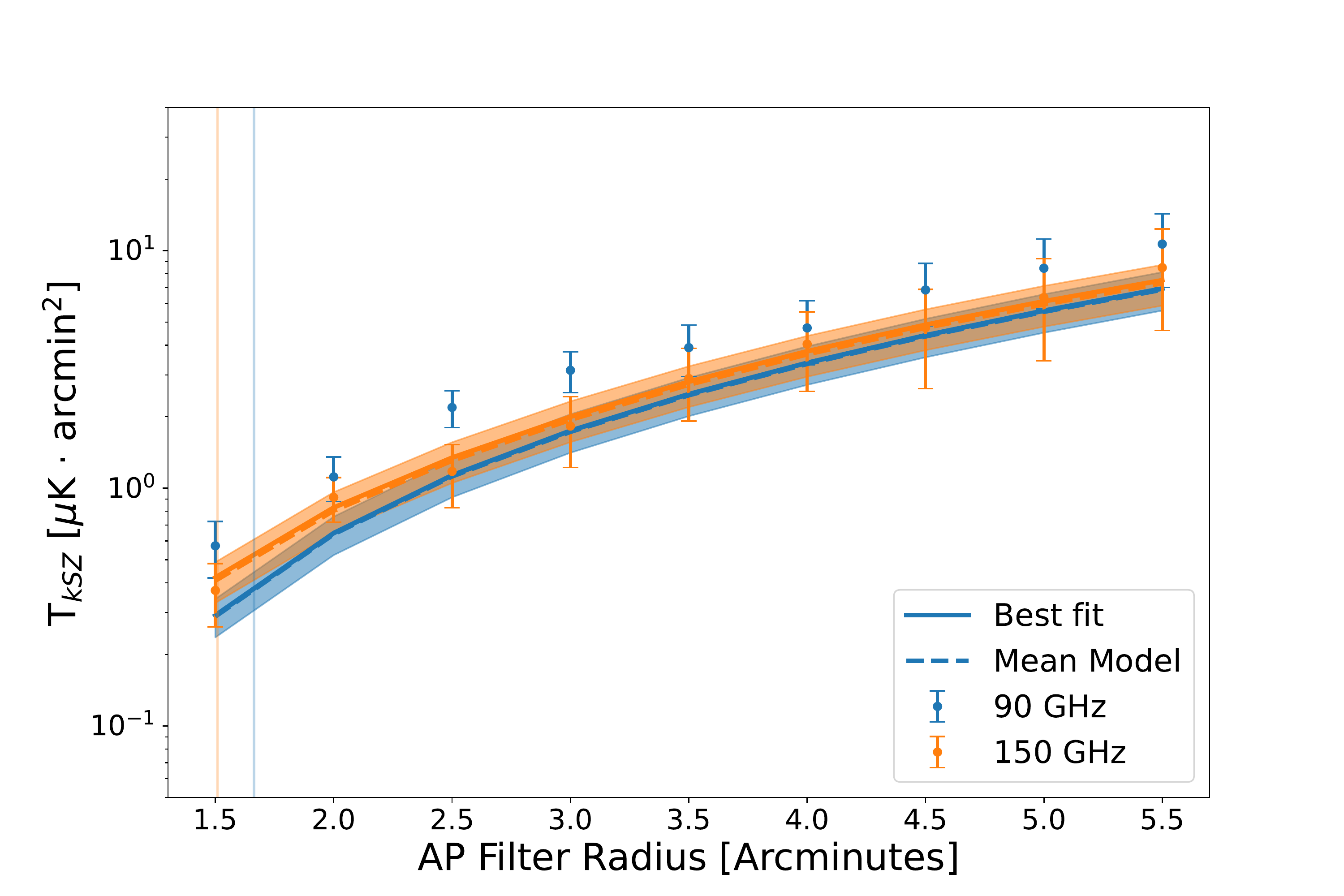}
    \caption{The joint fit to the f090 and f150 data. The band represents the 1$\sigma$ posterior spread of the fits. The best-fit model given by the solid lines is detected at the 4.8$\sigma$ level. We find that the model expects a stronger signal at f150, while the data has a stronger signal at f090. This tension is interesting to consider and is discussed more throughout the paper (see Sec \ref{sec:combfreq} and \ref{sec: null}), however, because the f090 and f150 error bars overlap for all but one bin, the significance of the tension is minimal.}
    \label{fig:joint_ksz}
\end{figure}

We find that jointly fitting the two frequencies does not improve our overall detection SNR for the best-fit model (4.8$\sigma$ compared to 6.2$\sigma$ and 4.6$\sigma$ for the f090 and f150 fits, respectively). Because this fit is less well constrained, we elect to fix our shape parameters ($\log_{10}\rho_0$\footnote{We note that $\log_{10}\rho_0$ represents the amplitude of the radial density profile, because this profile is used to model a 3D gas density, projected onto a 2D surface and then filtered with AP filters it is not the same as the amplitude factor $s$ which is used to fit the amplitude of the final kSZ profile.}, $x_c$, $\beta_k$  and $\textrm{A}_{2H}$) to the best-fit values for the minimum $\chi^2$ solution found using \texttt{scipy optimize} and then fit an overall amplitude using \texttt{emcee}. Doing so gives us a 4.8$\sigma$ detection and tightens the constraint on the amplitude to $S= 0.7\pm0.1$. The best-fit values from the individual frequency and combined fits are given in Table \ref{table: gnfw - ksz} and the results of the combined model are shown in Figure \ref{fig:joint_ksz}.

In A21, the authors also fit the same GNFW profile to their measurement of the kSZ signal; we find that our results agree with theirs at the 2$\sigma$ level. They measured $\textrm{log}_{10}\rho_0 = 2.6^{+0.4}_{-0.3}$, $x_c = 0.6 \pm 0.3$, $\beta_k = 2.6^{+1.0}_{-0.6}$ and $\textrm{A}_{2H} = 1.1^{+0.8}_{-0.7}$. In comparison, we find that our model for the combined frequency fit favors a higher best-fit value for $\textrm{A}_{2H}$, a marginally higher best-fit value for $x_c$, and a slightly lower best-fit value for $\textrm{log}_{10}\rho_0$. For the individual frequency fits, we find that our model again favors a higher best-fit value for $\textrm{A}_{2H}$. The values in A21 were calculated for a portion of the BOSS CMASS halos; that sample has a lower average redshift and a larger average mass than the DES redMaGiC halos studied here.  In general, halos of different mass and redshift are likely to have different gas profiles, due to differing impacts of feedback and structure growth; while these two samples' properties are currently consistent with one another, that may not be the case in future, higher-precision measurements.

Finally, we consider the reduced statistical power of the joint fit compared to the single band fits, which appears to arise from the differing central values of the single-frequency fits.  Figure \ref{fig:joint_ksz} shows that the f090 data points tend to lie above the f150 data, while the model expects these to be identical except for beam convolution, which would lead to a small difference in the opposite direction. The observed difference in signal between the two frequency channels makes it impossible for the GNFW profile to perfectly describe our measurements and reduces the overall detection significance in the combined fit.

The error bars on the f090 and f150 data are such that this difference could easily be caused by 
a statistical fluctuation and not be an indication of any true tension beyond noise. Alternatively, this tension could point to a systematic error that we have not been able to identify. Such a systematic error could also be the reason for the marginally passing null test between the f090 and f150 data described in Section \ref{sec: null}.

\subsection{Modeling the tSZ signal}
For the tSZ profile, we fix $\gamma_t = -0.3$ and we set $x_{c,t}$ using the results from \cite{2012batt}.

We then assume uniform priors of: 0.1 $< P_0 <$ 30, 0.1 $< \alpha_t <$2, 1 $< \beta_t <$ 10, and 0 $< \textrm{A}_{P2H}<$ 5.

\begin{figure}
    \centering
    \includegraphics[width = \linewidth, trim = 20cm 2cm 12cm 1cm, clip]{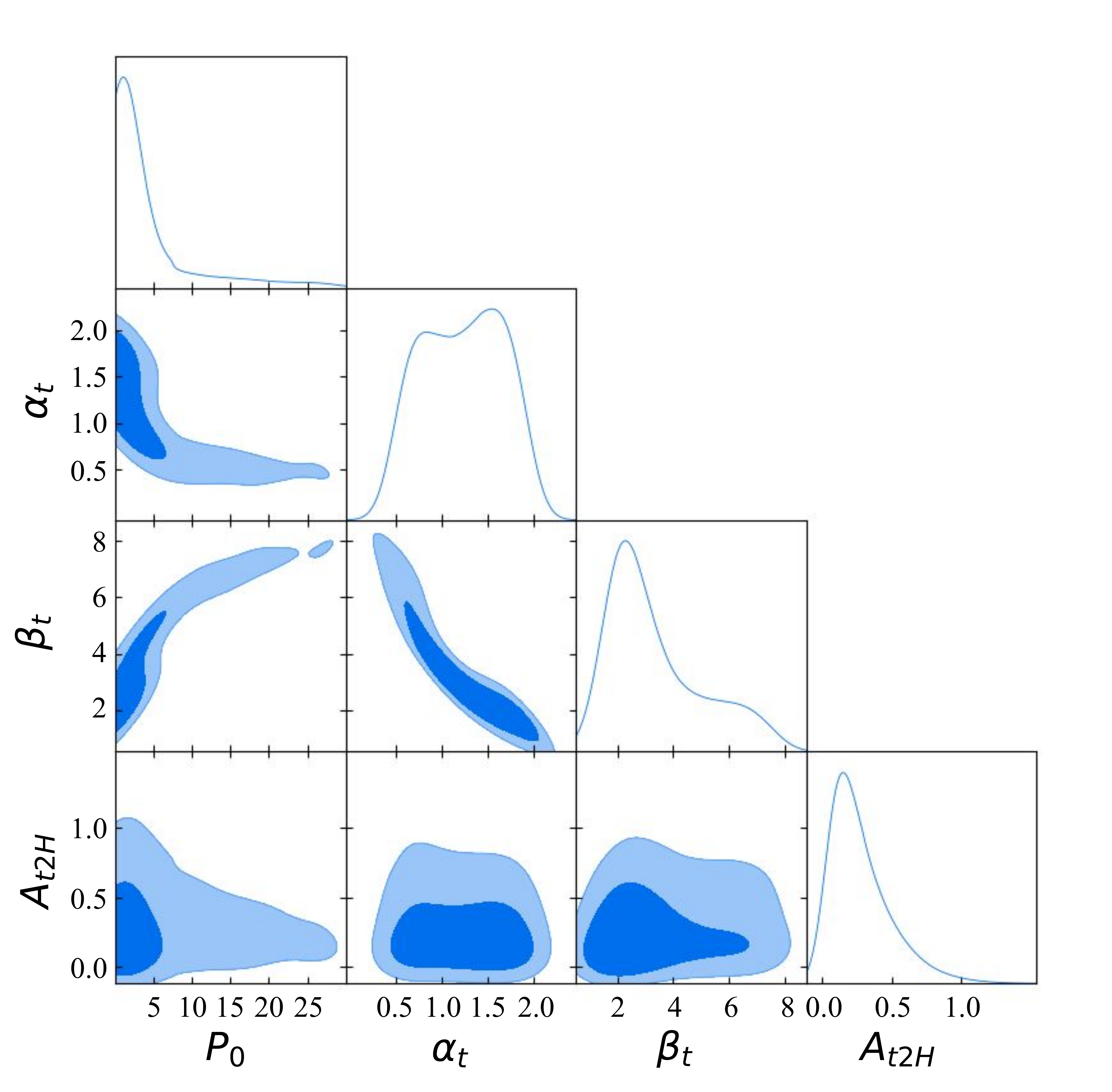}
    \caption{The GNFW pressure profile posteriors.}
    \label{fig: likelihood - tsz}
\end{figure}

\begin{figure}
    \centering
    \includegraphics[clip, trim=8cm 1cm 0cm 1cm, width = \linewidth]{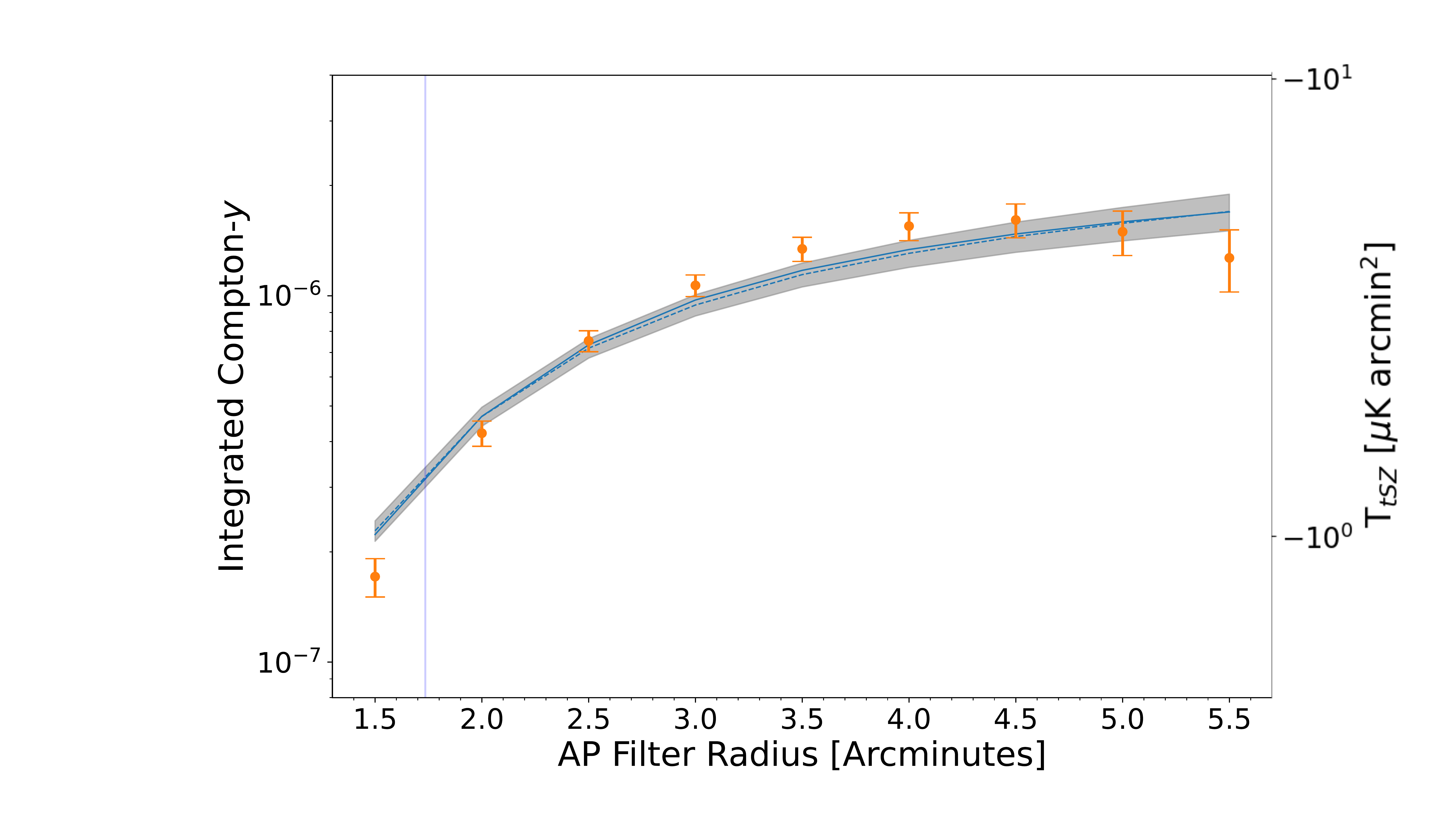}
    \caption{GNFW model of the tSZ profile. The gray band gives the $1\sigma$ spread of the posterior, while the dashed blue line represents the 50th percentile and the solid blue curve gives the maximum-likelihood solution. We plot the profile in Compton-$y$ units on the left-hand side, and on the right-hand side we give the signal in temperature units at 150 GHz. We also plot the virial radius, added in quadrature with the beam, as a vertical blue line. We measure the SNR of the best-fit model to be $\sqrt{\chi^2_{\rm null} - \chi^2_{\rm best-fit}} = 16.2$.}
    \label{fig: model_fit - tsz}
\end{figure}

The results for the tSZ parameter constraints are given in Table \ref{table: gnfw - tsz} and Figure \ref{fig: likelihood - tsz}. The model is then plotted in Figure \ref{fig: model_fit - tsz}.

We find that our model is favored over null at 16.2$\sigma$. This higher detection is due mainly to the fact that the tSZ signal is much stronger than the kSZ signal, which in turn makes it much easier to fit our model. While the model provides a good fit to the data we do see some evidence for an increase in the signal (in units of $\mu$K arcmin$^2$) at large radii. This increase is not yet statistically significant due to the large error bars at these radii, but would be interesting to study in more detail in the future.

As has been seen in previous tSZ studies, we note the degeneracy between $\beta_t$ and $P_0$. There is also a degeneracy between $\beta_t$ and $\alpha_t$ that is observable in the measurements from A21. We find that our fits are completely consistent with A21. The only deviation we see is for our measurement of the two-halo amplitude, for which they found a 1.8$\sigma$ preference for a non-zero two-halo term, but for which we find no evidence. However, they measured the tSZ signal using both the ILC maps (as done in this analysis) and single-frequency CMB maps combined with a dust model. Their fiducial analysis was based on the CMB maps + dust model and when compared to that analysis we see the difference mentioned above. However, if we compare to their measurements done using the same ILC maps used for this analysis, we find that this slight tension in the two-halo amplitude disappears. 

\begin{table}
\def\arraystretch{1.5}
\setlength\tabcolsep{5pt}
\begin{tabular}{|c|c|c|}
\hline
\thead{GNFW Pressure \\Parameters} & Priors & Constraints\\
\hline
$P_0$ & [0.1, 30] &$2.0^{+11.0}_{-1.0}$\\
$\alpha_t$ &[0.1, 2]& $0.9^{+0.6}_{-0.4}$\\
$\beta_t$ &[1, 10]&$4.0^{+3.0}_{-2.0}$ \\
$\textrm{A}_{P2H}$ &[0, 5]& $0.1\pm0.1$\\
\hline
\end{tabular}
\caption{The marginalized constraints from our MCMC for the tSZ profiles. \label{table: gnfw - tsz} }
\end{table}

We leave a more extensive analysis of the SZ signals to future work. These analyses could include comparing the results of these models to simulated galaxies. Such comparisons could result in improving our understanding of feedback mechanisms and galaxy evolution.

\subsection{Modeled Electron Temperature Profile \label{sec:elec_model}}

Given a model for the kSZ profile as well as one for the tSZ profile, it is also possible to extract an electron temperature profile from the two models as was done in A21. We use the same process, as outlined in Section \ref{sec:electron_temp}. The temperature profile is shown in Figure \ref{fig:etemp}. 

We find that the temperature profile generated by the models agrees well with the data, and all the data points lie within 1$\sigma$ of the model. We do notice a slightly higher temperature is preferred by the data over the model for the central bins with radii between 3\textquotesingle \
 and 4.5\textquotesingle. This effect is also seen in the Compton-$y$ data, which prefers a slightly higher value for Compton-$y$ than the model for these same bins.

Both the model and the data point to a temperature profile that decreases at higher radii, such that the temperature drops below the virial temperature beyond the virial radius. We caution that this profile is still an aperture photometry filtered profile, meaning it should not be interpreted as a radial temperature profile but instead can be thought of as an integrated profile. This suggests that the underlying radial temperature profile should drop off more steeply than the aperture photometry filtered profile shown here.

\section{\label{sec:discussion} Discussion}

In this paper, we present a new hybrid estimator that results in a 4.8$\sigma$ detection of the kSZ effect. We expect our SNR to scale with the square root of the sky area and for it to increase with more objects, although simulations would be needed to determine a more exact scaling. For this reason, we compare the SNR per object to a recent measurement done using exclusively spectroscopic data in S21. The comparison, shown in Table \ref{tab:detection_Sig}, demonstrates that, while our signal-to-noise is lower than that of the spectroscopic measurement, our method still performs well. This is an indication that this estimator will be an important tool for future high statistical significance kSZ studies.

\begin{table}[]
    \centering
    %\setlength\tabcolsep{5pt}
    %{\renewcommand{\arraystretch}{1.5}
    \begin{tabular}{|c|c|c|c|c|c|}
    \hline\\
        Paper & \thead{Sky area \\ deg$^2$} &\thead{Number of \\ Spectroscopic\\ Objects} & \thead{Number of \\ Photometric \\ Objects} & SNR & \thead{SNR per \\ 100,000 \\objects}  \\
        \hline
        This work & 1000 & 51,000 & 256,000 & 5.1$\sigma$ & 2.0$\sigma$\\
        S21 & 4000 & 311,000& 0 & 6.5$\sigma$ & 2.1$\sigma$\\
        \hline
    \end{tabular}
    \caption{Comparison of these results to those of S21, which used the same stacking method but with exclusively spectroscopic data. In both cases, we take the SNR for rejecting the null hypothesis (S21 also quotes an SNR for their best-fit model of 7.9$\sigma$). To calculate the SNR per 100,000 objects we consider just the objects that are used for the stacking analysis, for this work that is the 256,000 photometric objects and for S21 it is the 311,000 spectroscopic objects. We find that our SNR per object is lower than in S21, but still competitive when taking into account potential gains in the number of objects available for future photometric surveys. }
    \label{tab:detection_Sig}
\end{table}

In addition to the new hybrid kSZ estimator and the kSZ detection, we also measure the tSZ signal associated with these galaxies, the electron temperature profile and fit a GNFW profile to the SZ profiles. These measurements complement our main kSZ measurement and allow us to further probe the distribution of gas in these halos.  

We find evidence for an extended gas profile with mean temperature close to the virial temperature expected for halos of this size, in agreement with S21 and A21. Understanding these density profiles is a key ingredient to modeling galaxy feedback mechanisms, and the density profiles can be compared to simulations to determine whether simulations predict similar gas profiles to those detected via SZ measurements (see e.g. \cite{Kim_2022} for an example). This type of study was done in A21 where they found that at larger radii the measured pressure profiles are significantly higher than those predicted by simulations. Here, we expanded the work done by A21 by measuring and modeling density and pressure profiles for a different sample of galaxies. Because a comparison to simulations would depend on selecting a sample that is representative of the catalog used for this work, a precise, direct comparison between A21 and this work is not possible. A qualitative comparison indicates that the density profiles measured in this work are similar to those of A21.

Future studies will continue to improve upon these measurements and our ability to measure the gas profiles of high-redshift galaxies. Going forward, we expect the signal-to-noise of our kSZ measurements to scale as the square root of the sky area, or the square root of the number of objects times the mass of the objects for surveys with different object densities and mass distributions. Additionally, at small scales, we can expect to see some improvements as the depth of our CMB maps increase.

We can also consider the improvements we might see with future photometric and spectroscopic catalogs. Using LSST \citep{lsstsciencecollaboration2009lsst} and DESI \citep{desicollaboration2016desi} in the future would mean an increase in both depth and sky fraction. DESI projects they will observe 30 million galaxies, while LSST forecasts 20 billion. In terms of density, this means that DESI would observe $\sim$ 2000 objects per deg$^2$ while LSST will observe $\sim$ 650,000 objects per deg$^2$. Because photometric surveys often observe less massive objects, and the kSZ signal scales with mass, a direct comparison based on the number of objects is not accurate. However, due to the significantly higher density of the photometric catalogs, the inclusion of this data should still result in gains over using exclusively the spectroscopic data. In addition to the added depth, these surveys should have of order 6,000 deg$^2$ of overlap with Simons Observatory (SO) \citep{Ade_2019} and 7,000 deg$^2$ with CMB-S4 \citep{abazajian2016cmbs4, Pathways}. The hybrid estimator presented here is one method that could be used to take advantage of this data in the future.

The signal-to-noise of this measurement is also dependent on the depth of the CMB maps available to us. For this study, we have used ACT DR5 maps, which include data through 2018 from ACT. With subsequent seasons of data from ACT and future measurements from SO and CMB-S4 we expect to see improvements to the CMB maps that will translate to improvements for these kSZ measurements. 

We note that the methodology behind the velocity reconstruction for the spectroscopic data from BOSS is constantly evolving and improving (see e.g., \cite{Kitaura_2020}). These improved velocity reconstructions will also lead to improvements with this technique. 

\section{\label{sec:conclusion} Conclusion} 

High signal-to-noise kSZ measurements depend on having both high density galaxy catalogs and high-quality velocity reconstruction. Using our new hybrid kSZ estimator, we demonstrate that those elements can come from different datasets without having a substantially negative impact on the measurement. We also present a complementary measurement of the tSZ signal and constraints on the thermodynamics of the DES redMaGiC halos. Finally, we use the two SZ signals to extract an electron temperature profile.

Our hybrid kSZ estimator uses the spectroscopic catalog to estimate the underlying velocity field and then interpolates over that field to reconstruct the line-of-sight velocity data for the photometric catalog. From there, we apply AP filters to the ACT DR5 CMB maps at the location of the objects and weight the profiles by the reconstructed velocity. This pipeline leverages the more precise spectroscopic catalog while still taking advantage of the greater depth of the photometric catalog. While this technique does not yet result in the highest signal-to-noise, it does establish a new method that, when expanded to larger regions, should prove to be a valuable technique.

Going forward, this method will make it possible to leverage the incredible depth of photometric surveys to probe the gas profiles of galaxies in increasing detail, and will shed new light on the evolution of galaxies and galaxy clusters.

\section{\label{sec: acknowledgments} Acknowledgments}
ACT operated in the Parque Astron\'omico Atacama in northern Chile under the auspices of the Agencia Nacional
de Investigaci\'on y Desarrollo (ANID). Support
 for ACT was through the U.S.~National Science Foundation through awards AST-0408698, AST-0965625, and AST-1440226 for the ACT project, as well as awards PHY-0355328, PHY-0855887 and PHY-1214379. Funding was also provided by Princeton University, the University
 of Pennsylvania, and a Canada Foundation for Innovation (CFI) award to UBC.  The development of multichroic
 detectors and lenses was supported by NASA grants NNX13AE56G and NNX14AB58G. Detector research at NIST was supported by the NIST Innovations in Measurement Science program. Computing
 for ACT was performed using the Princeton Research Computing resources at Princeton University, the National Energy Research Scientific Computing Center (NERSC),  and the Niagara supercomputer at the SciNet HPC Consortium. 
 
MMK is
supported by the NSF Graduate Research Fellowship under Grant No. DGE-1256260. JCH acknowledges support from NSF grant AST-2108536, NASA grants 21-ATP21-0129 and 22-ADAP22-0145, DOE grant DE-SC00233966, the Sloan Foundation, and the Simons Foundation. CS acknowledges support from the Agencia Nacional de Investigaci\'on y Desarrollo (ANID) through FONDECYT grant no.\ 11191125 and BASAL project FB210003. This work was supported by a
grant from the Simons Foundation (CCA 918271, PBL). KM acknowledges support from the National Research Foundation of South Africa. EC acknowledges support from the European Research Council (ERC) under the European Union’s Horizon 2020 research and innovation programme (Grant agreement No. 849169).

We would also like to gratefully acknowledge the many publicly available software packages that made this work possible. We relied heavily on \texttt{emcee} \citep{Foreman_Mackey_2013}, \texttt{Mop-c-GT} \footnote{\url{https://github.com/samodeo/Mop-c-GT}} A21, \texttt{pixell} \footnote{\url{https://github.com/simonsobs/pixell}}, and \texttt{Astropy},\footnote{\url{http://www.astropy.org}} a community-developed core Python package for Astronomy \citep{astropy:2013, astropy:2018}. We also used \texttt{HEALPix} \citep{Gorski_2005} and the python wrapper for \texttt{HEALPix} which is \texttt{healpy} \citep{Healpix1}. In addition to these we made use of  \texttt{libsharp} \citep{reinecke/2013}, the \texttt{matplotlib} \citep{Hunter:2007} package, \texttt{numpy} \citep{oliphant2006guide, van2011numpy, harris2020array}, \texttt{pandas} \citep{jeff_reback_2020_3715232, mckinney-proc-scipy-2010} and \texttt{scipy} \citep{2020SciPy-NMeth}.

\bibliographystyle{aasjournal.bst}
\bibliography{main.bib}
\appendix

\section{Correlation and Degrees of Freedom \label{appendix: cor}}
\begin{figure}
    \centering
    \includegraphics[width=\linewidth]{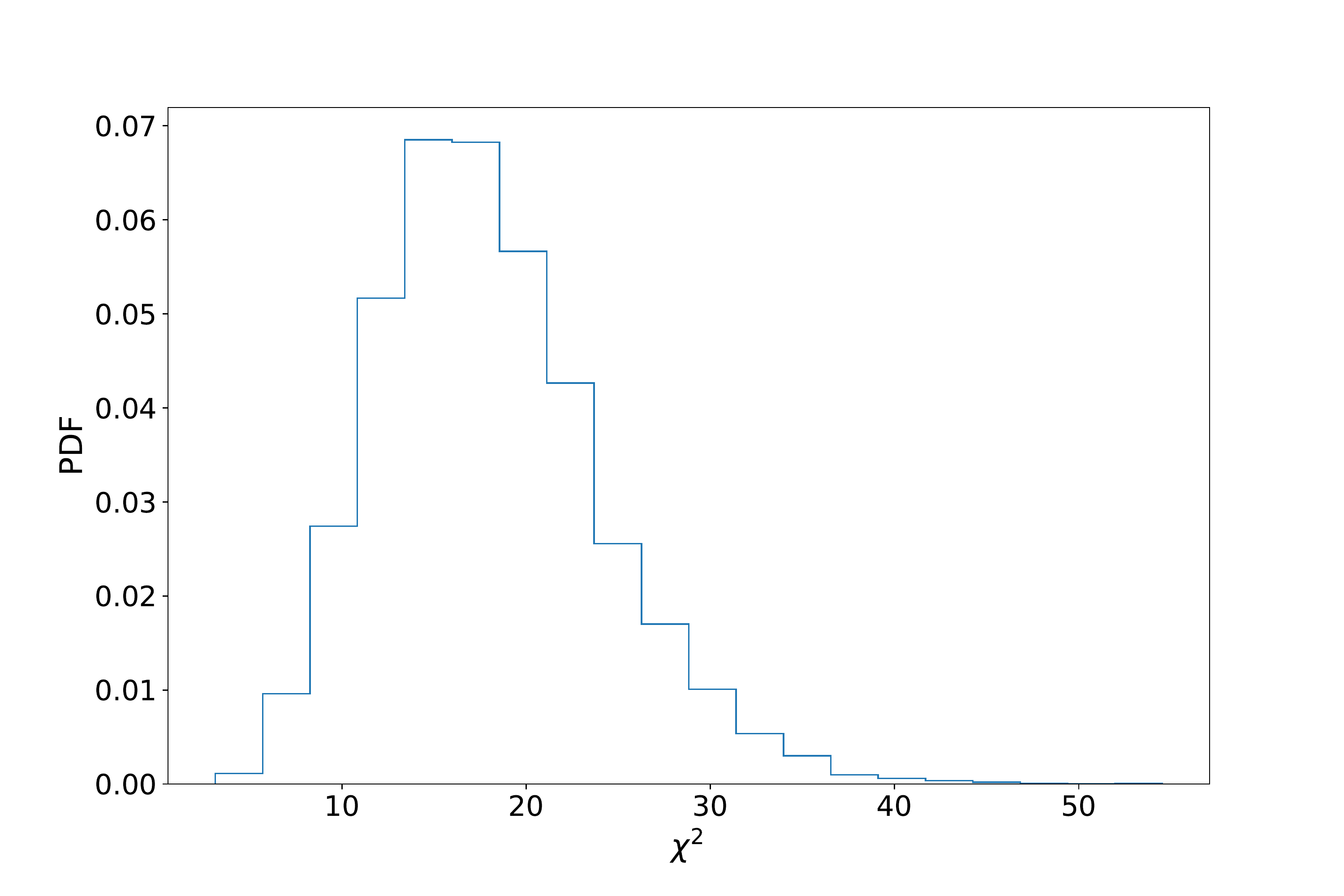}
    \caption{The probability density function of the simulated $\chi^2$ associated with the kSZ covariance matrix. The mean of these simulations is 18 and is consistent with the number of spatial degrees of freedom.}
    \label{fig: pdf}
\end{figure}

In order to convert the $\chi^2$ measurements in this paper to signal-to-noise measurements, we assume our kSZ measurement has 18 degrees of freedom, consistent with the number of spatial bins used. However, large correlations in the covariance matrix could mean that the true number of degrees of freedom is not consistent with the spatial degrees of freedom. This would then bias the PTEs for both the measurements and the null tests. Here, we examine this possibility by simulating the $\chi^2$ distribution associated with the covariance matrix. To do so, we repeatedly draw random samples from the multivariate normal distribution described by the covariance matrix. We repeat this process 10,000 times and calculate the $\chi^2$ for each sample. The result is shown in Figure \ref{fig: pdf}.  We find that the mean $\chi^2 = 18$, which is consistent with the number of spatial degrees of freedom.

\end{document}